\documentclass[11pt,a4paper]{article}
\usepackage{jheppub}

\usepackage{dcolumn}
\usepackage{latexsym}

\newcommand{\be}{\begin{equation}}
\newcommand{\ee}{\end{equation}}
\newcommand{\bea}{\begin{eqnarray}}
\newcommand{\eea}{\end{eqnarray}}

\newcommand{\refc}[1]{(\ref{#1})}
\newcommand{\Ord}{O}
\newcommand{\ew}[1]{\left\langle #1 \right\rangle}

\newcommand{\bt}{\bar{b}_2}
\newcommand{\btt}{$\bt$}

\newcommand{\SU}{\text{SU}}

\newcommand{\cont}{^{\rm cont}}
\newcommand{\eff}{^{\rm eff}}

\title{Spectrum of the open QCD flux tube and its effective string description
I: 3d static potential in SU({\boldmath $N=2,3$})}

\author{Bastian B. Brandt}
\affiliation{Institute for Theoretical Physics, Goethe University, \\
Max-von-Laue-Strasse 1, 60438 Frankfurt am Main, Germany}
\emailAdd{brandt@th.physik.uni-frankfurt.de}

\abstract{
We perform a high precision measurement of the static $q\bar{q}$ potential in
three-dimensional $\SU(N)$ gauge theory with $N=2,3$ and compare the results to
the potential obtained from the effective string theory. In particular, we
show that the exponent of the leading order correction in $1/R$ is 4, as predicted,
and obtain accurate results for the continuum limits of the string tension and the
non-universal boundary coefficient $\bar{b}_2$, including an extensive analysis
of all types of systematic uncertainties. We find that the magnitude of
$\bar{b}_2$ decreases with increasing $N$, leading to the possibility of a
vanishing $\bar{b}_2$ in the large $N$ limit. In the standard form of the
effective string theory possible massive modes and the presence of a rigidity
term are usually not considered, even though they might give a contribution to
the energy levels. To investigate the effect of these terms, we perform a
second analysis, including these contributions. We find that the associated
expression for the potential also provides a good description of the data. The
resulting continuum values for $\bar{b}_2$ are about a factor of 2 smaller than
in the standard analysis, due to contaminations from an additional $1/R^4$ term.
However, $\bar{b}_2$ shows a similar decrease in magnitude with increasing $N$.
In the course of this extended analysis we also obtain continuum results for the
masses appearing in the additional terms and we find that they are around twice
as large as the square root of the string tension in the continuum and compatible
between SU(2) and SU(3) gauge theory. In the follow up papers we will
extend our investigations to the large $N$ limit and excited states of the open
flux tube.
}


\begin{document}

\maketitle

\section{Introduction}

Flux tube formation between quark $q$ and antiquark $\bar{q}$ provides a
possible mechanism for quark confinement and has been verified in simulations
of lattice QCD (e.g.~\cite{Bali:1994de}). While the microscopic origin of the
formation of the flux tube is still debated, it is common consensus that the
long distance properties, in particular the spectrum, are well described by an
effective string theory (EST). The EST is a two dimensional effective field
theory for the Goldstone bosons associated with the breaking of translational
symmetry, the quantised transversal oscillation modes. While the basic
properties of the theory are known for a long
time~\cite{Goto:1971ce,Goddard:1973qh,Nambu:1978bd,Luscher:1980fr,
Polyakov:1980ca} a number of features have only been elucidated in the past
decade~\cite{Luscher:2004ib,Aharony:2009gg,Aharony:2010db,
Aharony:2010cx,Aharony:2011ga,Billo:2012da,Dubovsky:2012sh,Dubovsky:2012wk,
Aharony:2013ipa,Caselle:2013dra,Dubovsky:2014fma,Caselle:2014eka}.
In particular, it has been
clarified~\cite{Dubovsky:2012wk} that the leading order spectrum agrees with the
light cone quantisation~\cite{Arvis:1983fp} of the Nambu-Goto string theory (LC
spectrum, eq.~\refc{eq:LC-spectrum}) and the corrections to the LC spectrum
have been computed up to
$O(R^{-5})$~\cite{Aharony:2010db,Aharony:2011ga,Caselle:2013dra}, where $R$ is
the distance between quark and antiquark. For more details and a recent review
see~\cite{Brandt:2016xsp}.

The predictions of the EST can be compared to lattice results for the
excitation spectrum of the flux tube and good agreement has typically been
found, proceeding down to $q\bar{q}$ separations where the EST is not expected
to be reliable (for a compilation of results see~\cite{Brandt:2016xsp}). In
particular, the EST predicts a boundary correction of $O(R^{-4})$ with a free
and, presumably, non-universal coefficient \btt. This coefficient has first been
extracted from the excited states in 3d SU(2)~\cite{Brandt:2010bw} and later
from the groundstate in 3d $Z_2$~\cite{Billo:2012da} and
SU($N=2,3$)~\cite{Brandt:2013eua} gauge theory for a number of different lattice
spacings. Indeed, strong evidence for a non-universal behaviour has been found.
The good agreement down to small values of $R$ is surprising, given that the
flux tube profile only agrees with the EST starting from around 1.0~fm, at
least~\cite{Bali:1994de,Gliozzi:2010zv,Cardoso:2013lla,Caselle:2016mqu}. 
A possible explanation could be that the flux tube consists of a solid,
vortex-like core whose fluctuations are governed by the EST. This would explain
the good agreement of the profile with the associated exponential decay,
e.g.~\cite{Cea:2012qw,Cea:2014uja,Caselle:2016mqu}, rather then with the
Gaussian profile predicted by the EST~\cite{Luscher:1980iy,Caselle:1995fh},
while leaving the spectrum untouched to leading order.~\footnote{This can be
seen from the effective EST action which derives from vortex solutions of
the underlying microscopic theory~\cite{Forster:1974ga,
Gervais:1974db,Lee:1993ty,Orland:1994qt,Sato:1994vz,Akhmedov:1995mw}.} In
fact, it has been found~\cite{Cardoso:2013lla} that the profile can be
analysed using a convolution of the EST and vortex profiles, which would
indicate that the flux tube shares features of both.

One can expect that a particular class of corrections to the EST might show
up as massive
modes on the worldsheet. Indeed, candidates for states receiving
contributions from massive modes have been seen in 4d $\SU(N)$ gauge
theories~\cite{Morningstar:1998da,Juge:2002br,Juge:2004xr,Athenodorou:2010cs}
and the results for closed strings~\cite{Athenodorou:2010cs} are in good
agreement with the energy levels obtained by including a massive pseudoscalar
particle on the worldsheet known as the worldsheet
axion~\cite{Dubovsky:2013gi,Dubovsky:2014fma}. Recently, the behaviour of the
mass of the worldsheet axion for $N\to\infty$ has been
investigated~\cite{Athenodorou:2017cmw} and a finite result has been found
in this limit. It is interesting to note that possible massive modes are only
seen in 4d, which might be due to the fact that the topological coupling term
associated with the worldsheet axion only exists for $d>3$
(see section~\ref{sec:beyondEST}). Consequentially, any massive modes in
3d can only couple via terms contributing
to higher orders in the derivative expansion or via vertices including
more than one massive field. In this case results at intermediate
distances are potentially less affected and the massive modes appear as
quasi-free modes on the worldsheet.

Apart from the contribution of massive modes, there could also be contributions
from rigidity. The associated term satisfies the symmetry constraints of the
theory and naturally turns up in cases where the EST can be derived
from a vortex solution of the underlying field theory~\cite{Forster:1974ga,
Gervais:1974db,Lee:1993ty,Orland:1994qt,Sato:1994vz,Akhmedov:1995mw}.
While the contributions due to this term start at
higher orders in the derivative expansion of the EST (and thus should be
negligible up to $O(R^{-7})$, at least), it has been found, using zeta function
regularisation, that its non-perturbative (in the sense of the $1/R$ expansion)
contribution cannot be expanded systematically in $R^{-1}$.
It thus leads to contamination at all orders which are, however, exponentially
suppressed for large values of
$R$~\cite{Polyakov:1986cs,German:1989vk,Ambjorn:2014rwa,Caselle:2014eka}.
In fact, the leading order contribution is formally equivalent
to the contribution of a free massive particle, as discussed in
section~\ref{sec:beyondEST}.

In this series of papers we will investigate in detail the agreement of the
spectrum of the open flux tube in $\SU(N)$ gauge theories with the predictions
from the EST. Particular emphasis lies on the reliable extraction of the
boundary coefficient \btt{} from the static potential and its continuum
extrapolation. In this context it is also important to control the higher order
corrections. For the case that the other corrections are regular corrections
that are part of the derivative expansion, the next term would be of
$O(R^{-\gamma})$ with $\gamma\geq6$. Such corrections are comparably easy to
disentangle from the boundary term. If, on the other hand, there are
contributions from the rigidity term, or equivalently from massive modes, those
can contaminate the extraction of
\btt~\cite{German:1989vk,Ambjorn:2014rwa,Caselle:2014eka}. We will see
that we cannot exclude the latter possibility, but we can show that in a
certain range of distances the exponent of the correction term with respect to
the LC spectrum is indeed 4 and it is not well described by the additional terms
alone. Since we cannot exclude the contamination of \btt{}, we carry out two
independent analyses, excluding and including the rigidity/massive mode
contributions. Eventually we are interested in the large $N$ limit of the
non-universal contributions. In particular, there is hope that some of the
non-universal parameters can be used to constrain the possible holographic
backgrounds and eventually help to find the holographic dual of large $N$
Yang-Mills theory.

In this first paper of the series we will introduce the methods that we use to
analyse the potential in $\SU(N)$ gauge theory. However, we focus only on results
for $N=2$ and 3 in three dimensions and leave the extraction of the results for
$N>3$ and the extrapolation $N\to\infty$ for the next paper in the series. In
follow up papers we plan to consider excited states, extending the studies
from~\cite{Brandt:2009tc,Brandt:2010bw}, and
the four dimensional theory. The paper is organised as follows: In the next
section we briefly discuss the relevant predictions and properties of the EST and
its limitations. The consequent section~\ref{sec:groundstate} is devoted to
the basic analysis of the lattice data for the static potential. In particular,
we perform a reliable extraction of the Sommer parameter and the string tension
and show the leading correction to the LC spectrum is indeed of $O(R^{-4})$. In
sections~\ref{sec:b2-ana1} and~\ref{sec:rigidity} we extract
\btt{} excluding/including massive mode contributions,
respectively. Section~\ref{sec:results} provides the summary and discussion of
the main results of the two analyses and we conclude in 
section~\ref{sec:concl}. The details of the lattice simulations are presented in
appendix~\ref{app:sim-setup} and we discuss the control of systematic
uncertainties in appendix~\ref{app:sys-effects}.

\section{Effective string theory predictions}

We start by discussing the properties of the EST, focusing on the relevant 
aspects for the present study. For more detailed reviews we refer
to~\cite{Aharony:2013ipa,Brandt:2016xsp}.

\subsection{Effective string theory and its limitations}
\label{sec:est-setup}

The EST is the low energy effective field theory (EFT) describing a single,
stable, non-interacting flux tube. Here we will focus on the case of a flux
tube stretched between two sources in the fundamental representation
(quark and antiquark), but one can also study the hypothetical case of a
closed flux tube wrapping around a compactified dimension. The presence
of the flux tube breaks translational invariance in the transverse
directions, leading to $d-2$ Goldstone bosons (GBs), the quantised transverse
oscillation modes. For brevity we will consider the string of length $R$ to
reside in the $(x^0,x^1)$-plane ($x^0$ being the temporal direction) with
endpoints located at $x^1=0$ and $R$. Then one can parametrise the fluctuation
field by $X^\mu(x^\alpha)=(x^\alpha,\,X^i(x^\alpha))$, where $\alpha=0,1$,
$i=2,\ldots,d$ and the $X^i$ are the GBs of the EST. The action up to 6
derivatives order has been derived
in~\cite{Aharony:2009gg,Aharony:2010cx,Billo:2012da} together with the
constraints for the coefficients (the result for the coefficient $c_4$ has been
clarified in~\cite{Dubovsky:2012wk}). Here we will write down the action in the
static gauge (instead of the diffeomorphism invariant form discussed
in~\cite{Aharony:2013ipa,Brandt:2016xsp}) and in Minkowski space for simplicity.
Since we are considering an open string the action consists of two parts,
\be
\label{eq:full_action}
S_{\rm EST} = S_{\rm c} + S_{\rm b} \,,
\ee
where $S_{\rm c}$ is the bulk action,
\be
\label{eq:bulk_action}
\begin{array}{rcl}
\displaystyle S_{\rm c} & \displaystyle = \int_\mathcal{M} d^2 x & \displaystyle
\big[
-\sigma -\frac{\sigma}{2}\,\partial_\alpha X^i \partial^\alpha X^i + c_2
(\partial_\alpha X^i \partial^\alpha X^i)^2 + c_3 (\partial_\alpha X^i
\partial_\beta X^i)^2 \vspace*{2mm} \\ 
 & & \displaystyle \:\:\: + c_4
(\partial_\alpha\partial_\beta X^i \partial^\alpha \partial^\beta X^i)
(\partial_\gamma X^j\partial^\gamma X^j) +
c_5 (\partial_\alpha X^i\partial^\alpha X^i)^3 \vspace*{2mm} \\
 & & \displaystyle \:\:\: +c_6(\partial_\alpha X^i\partial^\alpha X^i)
(\partial_\beta X^j \partial_\gamma X^j)^2 + \dots \big] ,
\end{array}
\ee
where the integral runs over the worldsheet $\mathcal{M}$ of the flux tube, and
$S_{\rm b}$ the boundary action
\be
\label{eq:bound_action}
S_{\rm b} = \int_{\partial \mathcal{M}} d^2 x \:\big[ \mu + b_1 \partial_1 X^i
\partial_1 X^i\! + b_2 \partial_0\partial_1 X^i \partial_0\partial_1 X^i +
b_3(\partial_1 X^i\partial_1 X^i)^2\dots \big] ,
\ee
where the integral runs over the woldsheet boundary $\partial \mathcal{M}$.

The coefficients in eqs.~\refc{eq:bulk_action} and~\refc{eq:bound_action} are
constrained by the residual symmetries of the effective field theory, which for
the EST is Lorentz symmetry. The results for the coefficients
are~\cite{Aharony:2009gg,Aharony:2010cx,Billo:2012da,Dubovsky:2012wk}
\be
\label{eq:coeff_constr}
c_2=-\frac{\sigma}{8}\,,\quad c_3=\frac{\sigma}{4}\,,\quad
c_4=0\,,\quad c_5=\frac{\sigma}{16}\,,\quad
c_6=-\frac{\sigma}{8}\,,
\quad b_1=0\quad\text{and}\quad b_3=0 \,,
\ee
while $b_2$ remains unconstrained and thus represents the first subleading low
energy constant not fixed in terms of $\sigma$. In the next section we will
discuss the spectrum which follows from the effective
action. Note, that $b_2$ is a dimensionfull quantity, so that, for the purpose
of lattice simulations and the associated continuum limit, it makes sense to
introduce the dimensionless coupling
\be
\label{eq:bt-coupling}
\bt=\sqrt{\sigma^3}b_2 \,,
\ee
which we will use from now on.

On top of the terms in eq.~\refc{eq:full_action}, there is one more class of
terms allowed within the EST, constructed from powers of the extrinsic
curvature~\cite{Aharony:2013ipa,Caselle:2014eka}. The leading order term is
known as the rigidity term and has first been proposed by
Polyakov~\cite{Polyakov:1996nc}. It also turns up in those cases where vortex
solutions could be constructed from the underlying microscopic
theory~\cite{Forster:1974ga,Gervais:1974db,Lee:1993ty,Orland:1994qt,Sato:1994vz,
Akhmedov:1995mw}. In this sense, the presence of rigidity terms can be seen as
evidence of vortex contributions to the EST energies. Rigidity contributions to
energy levels start at higher orders in the perturbative expansion, but
it has been found that it leads to a non-perturbative contributions to the
energy
levels in zeta-function
regularisation~\cite{Braaten:1986bz,German:1989tr,Caselle:2014eka}. However, as
pointed out in~\cite{Dubovsky:2012sh}, this regularisation scheme does not
preserve the non-linear Lorentz symmetry of the EST, leading to possible
counterterms that need to be taken into account. It is thus mandatory to
crosscheck this result using a regularisation scheme which preserves Lorentz
invariance. In this paper we will use the result as presented
in~\cite{Caselle:2014eka}, which we summarise and discuss in more detail in
section~\ref{sec:beyondEST}.

The EST is expected to break down at the point where the energy of the
fluctuation modes reaches the QCD scale $\Lambda_{\rm QCD}\approx
\sqrt{\sigma}$. The energy of the modes is of the order $1/R$, meaning that the
EST is expected to break down for
\be
\label{eq:break-scale}
\sqrt{\sigma}R\lesssim 1 \,.
\ee
Owing to the derivative expansion in eqs.~\refc{eq:bulk_action}
and~\refc{eq:bound_action} this makes sense, since each derivative is of the
order of one unit of momentum of the degrees of freedom, $\partial\sim
p\sim1/R$, so that the EST corresponds to an expansion in
$(\sqrt{\sigma}R)^{-1}$ which is expected to break down when
eq.~\refc{eq:break-scale} is fulfilled. On top of this, there are also
several processes that are allowed in the microscopic theory, but are not
accounted for within the EST. Among them is the emission of glueballs. If those
are on-shell, the emitting state has to be an excited state with $E\geq E'+m_{G}$,
where $E'$ is the state it decays to and $m_G$ is the mass of the lightest
glueball with appropriate quantum numbers. Consequently, such a process can only
appear for excited states with $E_n>E_0+m_{G}$, where $E_0$ is the groundstate.
On top of this on-shell
emission, there can also be virtual glueball exchange. This process will
always be present (at finite $N$) and leads to corrections of unknown size.
These meson-glueball interactions and mixings are suppressed with increasing $N$
and vanish in the $N=\infty$ limit (e.g.~\cite{Lucini:2012gg}). Furthermore,
within the EST the string is not allowed to develop knots or to intersect with
itself, which would correspond to handles on the worldsheet. In addition, the
flux tube is likely to have an intrinsic width, allowing for inner excitations,
contributing in the form of massive excitations on the worldsheet which have to
be added to the EST. We will discuss these additional modes and their possible
contributions in section~\ref{sec:beyondEST}.

In general, the EST only describes a single non-interacting and stable string.
We would like to emphasise that this is not the case realised for the flux tube
in QCD, which can break due to the creation of a light $q\bar{q}$ pair from the
vacuum. Furthermore, the quarks are taken to be static, meaning that one takes
the limit of infinitely heavy quarks. The effect of finite quark masses can be
included using non-relativistic effective field theories of
QCD~\cite{Brambilla:2000gk,Pineda:2000sz,Brambilla:2003mu,Brambilla:2004jw}, for
instance (see~\cite{Brambilla:2014eaa} for a recent evaluation of the $1/m^2$
corrections to the potential using the EST predictions). In this paper, however,
we will focus on pure gauge theory, where effects owing to finite quark masses
are absent and the EST holds up to the limitations discussed above.

We close this section by noting that one possibility for an extension of the
standard EST to include the effect of possible internal excitation or external
influences on the flux tube is to view them as `defects' on the
worldsheet~\cite{Andreev:2012mc,Andreev:2012hw}. Here a defect is a place on the
string where the derivatives of the coordinates are discontinuous. The
associated effect has been worked out for the potential and found to be in
reasonable agreement with observed anomalous states in four
dimensions~\cite{Andreev:2012hw}.

\subsection{Spectrum of the open string and boundary corrections}
\label{sec:est-pred}

We will now discuss the spectrum that follows from the
action~\refc{eq:full_action}. The constraints in eq.~\refc{eq:coeff_constr} for
the coefficients set $c_2$ to $c_6$ to the values they obtain in the NG theory.
Consequently, one can expect that the leading order spectrum is equivalent to
the NG one, up to the corrections due to the boundary term proportional to
\btt. This expectation is corroborated by the formulation of the EST in
diffeomorphism invariant form~\cite{Aharony:2013ipa}, where the full NG action
appears as the leading order term in the action. The main question thus concerns
the spectrum following from the NG action in $d$ dimensions. For $d=26$ and
$d=3$ the exact spectrum is known from the light cone (LC) quantisation and
takes the form~\cite{Arvis:1983fp}
\be
\label{eq:LC-spectrum}
E^{\rm LC}_{n}(R) = \sigma \: R \: \sqrt{ 1 + \frac{2\pi}{\sigma\:R^{2}} \:
\left( n - \frac{1}{24} \: ( d - 2 ) \right) } \;.
\ee
Note, that we only consider open strings, so that $n$ denotes the number of
phonon excitations and as such labels the excitation level. The first few
excited states in terms of phonon creation operators $\alpha^i_{-m}$ (for phonon
momentum $m$) are listed in table~\ref{tab:3d-string-states}. However, the LC
quantisation breaks Lorentz invariance explicitly so that, away from $d=26$
and $d=3$, counterterms are necessary for a consistent quantisation. The first
counterterm is proportional to the $c_4$ term in eq.~\refc{eq:bulk_action} with
a coefficient~\cite{Dubovsky:2012sh,Dubovsky:2014fma}
\be
\label{eq:c4-counter}
\tilde{c}_4 = - \frac{d-26}{192\pi} \,.
\ee
Thus, the first correction in the bulk action to the LC spectrum appears at
order $R^{-5}$, but, since the NG action only contains one free parameter, the
correction is universal.

\begin{table}
\centering
\begin{tabular}{cc|cl|cc}
 \hline
 \hline
 energy & $\vert n, l \big\rangle$ & \multicolumn{2}{c|}{representation} &
$B_n^l$ & $C_n^l$ \\
 \hline
 \hline
 $E_0$ & $\vert0\big\rangle$ & $\mathtt{1} \vert0\big\rangle$ & scalar & 0 & 0
\\
 \hline
 $E_1$ & $\vert1\big\rangle$ & $\alpha^i_{-1} \vert0\big\rangle$ & vector & 4 &
$d-3$ \\
 \hline
 $E_{2,1}$ & $\vert2,1\big\rangle$ & $\alpha^i_{-1} \alpha^i_{-1}
\vert0\big\rangle$ & scalar & 8 & 0 \\
 $E_{2,2}$ & $\vert2,2\big\rangle$ & $\alpha^i_{-2} \vert0\big\rangle$ & vector
& 32 & $16 (d-3)$ \\
 $E_{2,3}$ & $\vert2,3\big\rangle$ & $\big(\alpha^i_{-1}
\alpha^j_{-1} - \frac{\delta^{ij}}{d-2} \alpha^i_{-1}
\alpha^i_{-1} \big) \vert0\big\rangle$ & sym. tracel. tensor & 8 & $4 (d-2)$\\
 \hline
 \hline
\end{tabular}
\caption{String states in the lowest three energy levels and their
representation in terms of creation operators $\alpha_m$ in the Fock space
together with their representation with respect to SO($d-2$). We also list the
associated values for the numbers $B_n^l$ and $C_n^l$ appearing in
eq.~\refc{eq:est-spec}. `sym. tracel.' for the state $\vert2,3\big\rangle$
stands for `symmetric traceless'.}
\label{tab:3d-string-states}
\end{table}

One might thus wonder whether the square root formula in
eq.~\refc{eq:LC-spectrum} should be taken as the starting point of the
expansion, or whether it is rather its expansion in orders of $R^{-1}$. In
fact, this has been one of the biggest puzzles associated with the EST in the
past decades. Lattice data (see~\cite{Brandt:2016xsp} for a compilation) show
excellent agreement with the full square root formula even down to values of
$R$ where its expansion breaks down. It thus has been conjectured that it
should be the full formula which provides a reasonable starting point
(e.g.~\cite{Athenodorou:2010cs,Brandt:2010bw}). The discussion above agrees with
this conjecture in the sense that the LC spectrum can schematically be written
as~\cite{Dubovsky:2012sh}
\be
\label{eq:ng-spect}
E^{\rm LC}_{n}(R) = E^{\rm NG}_{n}(R) - {\rm counterterms} \,.
\ee
Consequently, it will always be the full square root formula which appears when
we solve the above equation for $E^{\rm NG}_{n}$. Furthermore, an analysis
using the machinery of the Thermodynamic Bethe Ansatz (TBA) shows, that the
leading order $S$-matrix is integrable, leading to energies given by the full
square root formula~\cite{Dubovsky:2012sh,Dubovsky:2014fma}. Note, that the
boundary term can also be included in this TBA analysis~\cite{Caselle:2013dra}.

Following the above discussion and the corrections to the LC spectrum computed
in~\cite{Aharony:2010db,Aharony:2011ga} the spectrum up to $O(R^{-5})$ is thus
given by
\be
\label{eq:est-spec}
E^{\rm EST}_{n,l}(R) = E^{\rm LC}_{n}(R) - \bt \frac{\pi^3}{\sqrt{\sigma^3} R^4}
\Big( B_n^l + \frac{d-2}{60} \Big) - \frac{\pi^3 (d-26)}{48 \sigma^2 R^5}
C_n^l + \Ord(R^{-\xi}) \,,
\ee
where we have inserted the dimensionless coupling \btt{} introduced in the
previous section. $B_n^l$ and $C_n^l$ are dimensionless coefficients tabulated
in table~\ref{tab:3d-string-states}. They depend on the representation of the
state with respect to rotations around the string axis, transformations of
$X^i$ with elements of SO($d-2$), and thus lift the degeneracies of the LC
spectrum. Note once more, that $C_n^l$ vanishes identically for $d=3$ since
the state $\vert2,3\big\rangle$ does not exist in this case. This can also be
seen from the associated term in the EST action, the $c_4$ term in
eq.~\refc{eq:bulk_action}, which is trivial for $d=3$. The next correction term
to the LC energies can be expected to appear with an exponent $\xi=6$ or
$\xi=7$, depending on whether the next correction originates from another
boundary term (which will generically be the case if the associated
coefficient does not vanish identically due to symmetry) or a bulk term.

\subsection{Beyond the standard EST: massive modes and rigidity term}
\label{sec:beyondEST}

As mentioned already in the introduction and in section~\ref{sec:est-setup},
there is also another class of terms allowed within the EST containing the
extrinsic curvature. The leading order term of this type is known as the
rigidity term, whose presence has first been found in~\cite{Polyakov:1986cs}.
In terms of the derivative expansion the contributions of the rigidity term
start at 8th derivative order, but the presence of this term in the action
can give a non-perturbative contribution to the
potential~\cite{Klassen:1990dx,Nesterenko:1997ku,Caselle:2014eka}.
Including the first two terms from the bulk action, eq.~\refc{eq:bulk_action},
together with the leading order contribution from the rigidity term
and evaluating the resulting Gaussian integral leads to a Euclidean potential
of the form (for the details see~\cite{Caselle:2014eka})
\be
\label{eq:pot-rdet}
V(R) = \lim_{T\to\infty}\Big\{ \sigma R + \frac{1}{2T}
\log\Big[ \det\big(-\Delta\big) \,
\det\Big(1-\frac{\Delta}{m^2}\Big) \Big] \Big\} ,
\ee
where T is a finite temporal extent of the spacetime, $\Delta$ is the 2d
Laplace operator, and we have introduced the mass
\be
\label{eq:mdef}
m=\sqrt{\sigma/2\alpha} \,.
\ee
The first determinant is the one resulting from a free massless boson field,
leading to the Coulombic term within the EST, first computed by
L\"uscher~\cite{Luscher:1980ac}, while the second is reminiscent of
the one of a free boson field with mass parameter $m$.

At this point it is important to stress that the contribution of the rigidity
term resembles that of a massive excitation on the world sheet in Euclidean
spacetime. However, the above result is only the leading order result, originating
from the Gaussian part of the rigidity action. Higher order contributions will
potentially spoil this equivalence. Nonetheless, at this order the two types of
contributions cannot be disentangled.

Using zeta-function regularisation, the determinant of the massive boson in
eq.~\refc{eq:pot-rdet} leads to a leading order term in the potential of the
form~\cite{Klassen:1990dx,Nesterenko:1997ku,Caselle:2014eka}
\be
\label{eq:pot-rigid}
V^{\rm rig}_0(R) = - \frac{m}{2\pi} \sum_{k=1}^{\infty} \frac{K_1(2kmR)}{k} \,,
\ee
which appears on top of the leading order Coulomb term originating from the
massless determinant. Note, that the result has been obtained in
zeta-function regularisation which breaks Lorentz invariance explicitly. Owing
to the discussion in section~\ref{sec:est-setup} this means that one has to
care for the inclusion of potential higher order counterterms in the action
to obtain the correct result. The higher order terms in the action can be
included perturbatively~\cite{German:1989vk,Caselle:2014eka}, leading to
the additional term~\cite{Braaten:1986bz,German:1989vk}
\be
\label{eq:pot-rigid-cor}
V^{\rm rig}_1(R) = - \frac{(d-2)(d-10)\pi^2}{3840 m \sigma R^4} \,,
\ee
which contaminates the $R^{-4}$ boundary term from eq.~\refc{eq:est-spec}.
Naively we expect a similar term in the presence of a free boson on the
worldsheet.

To assess the effect of the term $V^{\rm rig}_0$ it is instructive to consider
the two different limits with respect to the EST. First, let us consider the
large $R$ limit, so that $mR\gg 1$. In this case the dominating contribution
comes from the term in the sum with $k=1$, which gives a leading order
contribution of the form~\cite{Caselle:2014eka}
\be
\label{eq:pot-rigid-largeR}
V^{\rm rig}_0(R) \approx -\sqrt{\frac{m}{16\pi R}} e^{-2mR} \,.
\ee
Consequently, $V^{\rm rig}_0$ will be exponentially suppressed with respect to
all other terms in the EST, so that it can only give a relevant contribution at
intermediate or small values of $R$ for a given mass $m$. Next, let us consider
the region of small $mR<\pi$. In this case one obtains~\cite{Caselle:2014eka}
(keeping only the terms relevant for $R\to0$)
\be
\label{eq:pot-rigid-smallR}
V^{\rm rig}_0(R) \approx -\frac{\pi}{24 R} + \frac{m^2 R}{4\pi}
\ln\Big(\frac{mR}{2\pi}\Big) + m\cdot O(mR) \,.
\ee
Thus, for small distances the rigidity term leads to another Coulombic term,
doubling the standard L\"uscher term within the EST. What is particularly
important is the fact that in some intermediate regime (which, for not too large
values of $m$, will still be within the validity region of $mR<\pi$
expansion) the Coulombic term will dominate over the $R^{-4}$ correction (while
the logarithm is already negative). Consequently, we expect a negative Coulombic
correction to the LC energy levels in some regime before seeing an $R^{-4}$
increase for $R\to0$ if the \btt{} term is absent.

As a remark, we would like to stress that the above computation only
applies to the potential. The modification of the excited energy levels due
to the rigidity term are still unknown. For the purpose of this paper we will
only need the potential, but a distinction between the boundary corrections and
the rigidity term is notoriously difficult in this case. It would be very
interesting to get a result for the associated modification of excited states.
The hope is, that for the excited states the predictions including the massive
modes is incompatible with the splittings between the energy levels, so that
one can distinguish between the cases with or without massive mode
contributions. In particular, it has been found in~\cite{Brandt:2010bw} that the
splitting of the first few excitations is well described by the boundary term,
which could be different when massive modes are included.

In the presence of additional massive bosonic degrees of freedom the presence
of terms proportional to the extrinsic curvature also allows for a
topological coupling term between the boson field and the GBs, proportional
to the mode number~\cite{Dubovsky:2013gi,Dubovsky:2014fma}~\footnote{Note, that
the presence of this worldsheet $\theta$-term has first been noticed
in~\cite{Polyakov:1986cs}.}. Consequently, since the boson couples to the
worldsheet analogue of the $\theta$-term, the boson can naturally be referred
to as the worldsheet axion~\cite{Dubovsky:2013gi}. The action for the
worldsheet axion can then be written as~\cite{Dubovsky:2014fma}
\begin{equation}
\label{eq:act-axion}
S_{\rm a} = \int_\mathcal{M}\sqrt{-h} \Big(-\frac{1}{2}\nabla_\alpha
\phi\nabla^\alpha \phi-\frac{1}{2}m^2\phi^2 -\frac{\gamma}{8\pi} \phi\,
\epsilon_{ij}\epsilon^{\alpha\beta}K_{\alpha\gamma}^iK_\beta^{\gamma i}+
\ldots\Big) ,
\end{equation}
where $\nabla$ is the covariant derivative with respect to the induced metric
$h^{\alpha\beta}=\partial^\alpha X \partial^\beta X$. To leading order, i.e.
up to coupling terms of the form $\phi\partial_\alpha\partial_\beta\phi
h^{\alpha\beta}$ and higher orders, $\nabla^\alpha$ can be replaced by
$\partial^\alpha$. The
associated leading order spectrum has been computed with the TBA method for
closed strings in~\cite{Dubovsky:2013gi,Dubovsky:2014fma} and has been found to
be consistent with states showing an anomalously slow approach to the LC
energies in the spectrum of the closed flux tube for
$d=4$~\cite{Athenodorou:2010cs}. Note that the topological coupling term in
eq.~\refc{eq:act-axion} is only present for $d>3$, which could be a reason why
no anomalous states have been observed in the 3d flux tube spectrum. In 3d
massive modes appear as free bosons up to the coupling terms mentioned above.
It would be interesting to include those coupling terms in a computation of the
energy levels to check their influence on the spectrum.

In the following we will always denote the contributions discussed in this
section as the contributions from ``massive modes'' for brevity. We would
like to emphasise, however, that strictly speaking we can only compare our
results to those obtained from the rigidity term, or from the leading order
contribution, neglecting any direct couplings, of a massive boson on the
worldsheet. Whether direct coupling terms are indeed negligible is an open
question, which we cannot answer in the course of this study.

\subsection{EST and gauge/gravity duality}

We close the discussion of the EST with the remark that the EST action can
potentially be computed (e.g.~\cite{Aharony:2009gg} and references therein) from
the original 10d fundamental string theory appearing in a generalisation of the
AdS/CFT correspondence for large $N$ gauge
theories~\cite{Maldacena:1997re,Gubser:1998bc,Witten:1998qj}. For recent
computations of properties of the flux tube in this framework
see~\cite{Kol:2010fq,Vyas:2010wg,Vyas:2012dg,Giataganas:2015yaa}. Consequently,
constraining the possible terms and their couplings in the EST action could
allow to constrain the fundamental string theory relevant for Yang-Mills theory.

In particular, in such cases where the background in the fundamental string
theory is weakly curved, the additional bosonic and fermionic degrees of freedom
(e.g. the additional coordinate fields in the directions transverse to the gauge
theory plane and the supersymmetry partners of the bosonic fields) can be
integrated out perturbatively. This has been done for closed strings and a
special class of backgrounds in~\cite{Aharony:2009gg} and the resulting terms
have been found to agree with the terms appearing in the EST bulk action,
eq.~\refc{eq:bulk_action}. Furthermore, in~\cite{Aharony:2010cx} it has been
shown that the same class of backgrounds considered for an open string ending on
two infinitely stretched $D$-branes leads to the boundary term proportional to
\btt{} in eq.~\refc{eq:bound_action}. In terms of the masses of the additional
bosons its contribution is given by~\cite{Aharony:2010cx}
\be
\label{eq:b2-holography}
b_2 = - \frac{1}{64 \sigma} \sum_\xi \frac{(-1)^{{\rm BC}(\xi)}}{m^b_\xi} +
b^f_2 + \ldots \,,
\ee
where $\xi$ labels the transverse directions to the gauge theory plane,
$m^b_\xi$ is the mass of the boson field associated with the direction $\xi$,
$b^f_2$ is the contribution from the fermionic degrees of freedom and
${\rm BC}(\xi)$ depends on the boundary conditions for the fields in this
direction. In the case of Dirichlet boundary conditions ${\rm BC}=0$, while for
Neumann boundary conditions one has ${\rm BC}=1$. The ellipses in
eq.~\refc{eq:b2-holography} stand for terms originating from possible other
fields present on the gravity side of the duality. We would like to emphasize
that, even though cancellations can appear, there is no reason why the terms in
eq.~\refc{eq:b2-holography} should add up to zero, except for certain fine-tuned
situations (see also the discussion in~\cite{Aharony:2010cx}), so that,
generically, one can expect \btt{} to be non-vanishing when the EST originates
from a duality with this types of string background. It would be interesting to
see whether the duality can also account for the rigidity term in the EST
action and make a statement about its coefficient.

The aforementioned computations have neglected
the appearance of possible additional massless scalar modes on the
worldsheet.~\footnote{Note, that those are present for all the examples
investigated in~\cite{Aharony:2009gg}.} If present, they would contribute to
the L\"uscher (Coulomb) term and thus inherently change the large $R$ behaviour
of the confining string. Since the L\"uscher term is well reproduced by
lattice data, this situation is basically ruled out. This issue can be resolved 
if the action for these additional degrees of freedom non-perturbatively develops
a mass gap, so that these additional fields contribute as massive degrees of
freedom on the worldsheet.

\section{Results for the potential}
\label{sec:groundstate}

\begin{table}
\centering
\begin{tabular}{cc|cccc|c}
 \hline
 \hline
 $\beta$ & Lattice & $R/a$ & $t_s$ & $n_t$ & \#meas & $T/r_0$ \\
 \hline
 \hline
 $\SU(2)$ & & & & & & \\
 5.0 & $48^3$ & 2-14 & 2 & 20000 & 1600 & 12.2 \\
 5.7 & $48^3$ & 1-13 & 2 & 20000 & 5000 & 10.4 \\
 6.0 & $48^3$ & 1-18 & 3 & 20000 & 7300 & 9.8 \\
 7.5 & $64^3$ & 1-24 & 4 & 20000 & 3000 & 10.2 \\
 10.0 & $96^3$ & 1-28 & 6 & 20000 & 3100 & 11.1 \\
 12.5 & $128^3$ & 1-34 & 8 & 20000 & 2400 & 11.7 \\
 16.0 & $192^3$ & 1-37 & 12 & 20000 & 2800 & 13.7 \\
 \hline
 $\SU(3)$ & & & & & & \\
 11.0 & $48^3$ & 2-11 & 2 & 20000 & 1700 & 14.6 \\
 14.0 & $48^3$ & 2-14 & 2 & 20000 & 1900 & 10.8 \\
 20.0 & $64^3$ & 1-23 & 4 & 20000 & 2700 & 9.5 \\
 25.0 & $96^3$ & 1-28 & 6 & 20000 & 2200 & 11.2 \\
 31.0 & $128^3$ & 1-33 & 8 & 20000 & 2400 & 11.8 \\
 \hline
 \hline
\end{tabular}
\caption{Simulation parameters for the extraction of the Polyakov loop
correlation functions. Listed are the range of $q\bar{q}$ separations, the
temporal extent of the LW sublattices $t_s$, both in units of the lattice
spacing, the number of sublattice updates $n_t$ and the number of measurements.
We also list the temporal lattice extent in units of the
Sommer scale $r_0$.}
\label{tab:sim-paras}
\end{table}

To extract the static potential we have used the spatial Polyakov loop
correlation function. The details of the simulations and the extraction of
the potential from the correlation function are discussed in
appendix~\ref{app:sim-setup}, where we also discuss the L\"uscher-Weisz (LW)
multilevel algorithm~\cite{Luscher:2001up}, which is mandatory to achieve
the precision needed for the extraction of the subleading coefficients
in the EST. For the suppression of excited state contaminations and finite size
effects our simulations have been done on large lattices. The parameters of the
simulations are tabulated in table~\ref{tab:sim-paras}. To demonstrate that,
even for the high precision achieved, the aforementioned effects are indeed
negligible, we report on extensive checks in appendix~\ref{app:sys-effects}.

\subsection{Scale setting and string tension}
\label{sec:scale-setting}

For scale setting purposes we use the Sommer parameter
$r_0$~\cite{Sommer:1993ce} (see appendix~\ref{app:sim-setup}). For $\SU(3)$ and
$d=4$ the associated continuum value is $r_0=0.5$~fm which can be used to
convert to ``physical units''. We extract $r_0/a$ from the force using four
different methods:
\begin{enumerate}
 \item[(a)] a numerical polynomial interpolation;
 \item[(b)] a numerical rational interpolation;
 \item[(c)] a parameterisation of the form~\cite{Sommer:1993ce}
 \be
  \label{eq:force-string}
  F(R) = f_0 + \frac{f_1}{R^2} + \frac{f_2}{R^4}
 \ee
 for the values of $R$ corresponding to the four nearest neighbours of $r_0$
 (motivated by the LO EST);
 \item[(d)] the parameterization of \refc{eq:force-string}
 with $f_2=0$ for the two nearest neighbours of $r_0$.
\end{enumerate}
As final estimate for $r_0/a$ we will use method (d). The systematic
uncertainty associated with the interpolation to obtain $r_0/a$ can be estimated
from the maximal deviation of $r_0/a$ obtained from methods (a), (b), and (c)
compared to method (d). The results for $r_0/a$ are tabulated in
table~\ref{tab:scale-fits}.

Another option for scale setting is to use the string tension $\sigma$,
associated with the $R\to\infty$ asymptotics of the potential. We extract the
string tension in two different ways:
\begin{enumerate}
 \item[(i)] we fit the data for the force to the form 
 \be
  \label{eq:force-sig-fit}
  R^2 F(R) = \sigma R^2 + \gamma \,,
 \ee
 following eq.~\refc{eq:force_nlo};
 \item[(ii)] we fit the potential to the leading order EST prediction,
 eq.~\refc{eq:LC-spectrum} for $n=0$, adding a normalisation constant $V_0$.
\end{enumerate}
Both ans\"atze are correct only up to a certain order in the $1/R$ expansion, so
that, in the region where higher orders become important, we will get incorrect
results for the string tension. To isolate the asymptotic linear behaviour
of the potential we investigate the dependence on the minimal value of $R$
included in the fit, $R_{\rm min}$. The strength of using these particular two
ans\"atze is, that the corrections to the fit formula are different, so that we
can determine the region where higher order terms are negligible by comparing
the results. In the region where both sets of results, in dependence on $R_{\rm
min}$, show the onset of a plateau and agree within errors, the estimate for
$\sigma$ will be reliable (with the present accuracy).

\begin{figure}[ht]
 \centering
\includegraphics[]{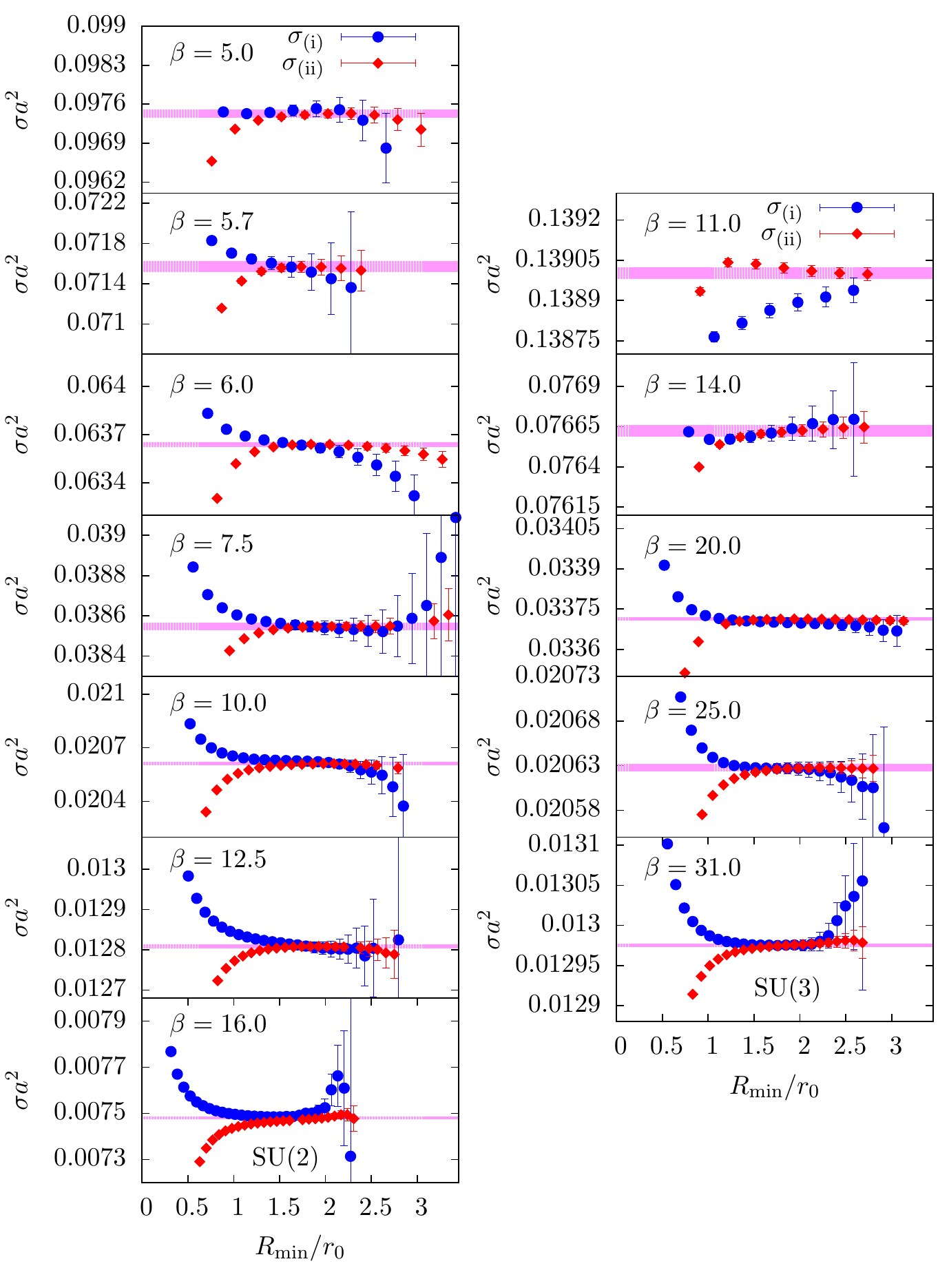}
 \caption{Results for the string tension in SU(2) (left) and SU(3) (right)
gauge theory extracted from method (i), $\sigma_{\rm (i)}$, and (ii),
$\sigma_{\rm (ii)}$, as explained in the text, versus the minimal value of $R$
included in the fit, $R_{\rm min}$ in units of $r_0$. The red bands are the
values for $\sigma_{(ii)}$ which we will use for the further analysis.}
 \label{fig:sigma_vs_Rmin}
\end{figure}

The results obtained from the two methods, denoted as $\sigma_{\rm (i)}$ and
$\sigma_{\rm (ii)}$, respectively, are shown in figure~\ref{fig:sigma_vs_Rmin}
for $\SU(2)$ (left) and $\SU(3)$ (right). The plots indicate that
in most of the cases the extraction of $\sigma$ is reliable. The most critical
case is $\beta=11.0$ for SU(3), whose value of $\sigma$ we exclude from the
following analysis. Another critical case is $\beta=16.0$ for SU(2), where the
two plateaus for $\sigma_{\rm (i)}$ and $\sigma_{\rm (ii)}$ do not agree within
errors. In that case we use the extent where the two results are closest (the
discrepancy is of order $1\sigma$). The resulting values for $\sigma$ are 
listed in table~\ref{tab:scale-fits} together with the other fit parameters and
are indicated by the bands in the figures. The plots indicate that within the
region where the two fits agree the particular choice for $R_{\rm min}$ does not
matter within the given uncertainties. In the following analysis we will use
$\sigma_{\rm (ii)}$, whose value is shown as the red band in
figure~\ref{fig:sigma_vs_Rmin}.

\begin{table}
 \centering
 \begin{tabular}{cc|r|rr|rr}
  \hline
  \hline
  & & & \multicolumn{2}{c|}{(i)} & \multicolumn{2}{c}{(ii)} \\
  $N$ & $\beta$ & \multicolumn{1}{c|}{$r_0/a$} &
  \multicolumn{1}{c}{$\sqrt{\sigma} r_0$}
  & \multicolumn{1}{c|}{$\gamma$} & \multicolumn{1}{c}{$\sqrt{\sigma} r_0$} 
  & \multicolumn{1}{c}{$aV_0$} \\
  \hline
  \hline
  2 & 5.0  & 3.9472(4)( 7)  & 1.2325(14)(1) & 0.129(16) & 1.2321(5)(1) &
  0.2148(6) \\
  & 5.7  & 4.6072(4)(10)  & 1.2321(16)(1) & 0.141(13) & 1.2325(4)(1) & 0.2031(4)
  \\
  & 6.0  & 4.8880(2)( 4)  & 1.2330( 3)(1) & 0.136( 2) & 1.2331(4)(1) &
  0.1976(1) \\
  & 7.5  & 6.2860(4)( 3)  & 1.2341( 6)(1) & 0.135( 6) & 1.2341(3)(1) & 0.1740(2)
  \\
  & 10.0 & 8.6021(4)( 8)  & 1.2349( 8)(1) & 0.136( 9) & 1.2350(3)(1) & 0.1449(2)
  \\
  & 12.5 & 10.9085(7)(16) & 1.2345( 7)(1) & 0.136( 6) & 1.2346(3)(1) & 0.1248(1)
  \\
  & 16.0 & 14.2958(6)(10) & 1.2401(32)(1) & 0.092(33) & 1.2364(5)(1) &
  0.1026(2)  \\
  \hline
  3 & 11.0 & 3.2881(2)( 6)  & 1.2256( 2)(1) & 0.139( 3) & 1.2259(1)(1) &
  0.2498(2) \\
  & 14.0 & 4.4433(3)( 1) & 1.2303( 9)(1) & 0.129( 9) & 1.2300(3)(1) & 0.2239(3)
  \\
  & 20.0 & 6.7075(4)( 2) & 1.2318( 2)(1) & 0.138( 2) & 1.2318(1)(1) & 0.18196(6)
  \\
  & 25.0 & 8.5797(4)( 6) & 1.2322( 3)(1) & 0.135( 3) & 1.2322(1)(1) & 0.15709(6)
  \\
  & 31.0 & 10.8182(4)( 5) & 1.2323( 3)(1) & 0.134( 3) & 1.2323(1)(1) &
  0.13550(5) \\
  \hline
  \hline
 \end{tabular}
\caption{Results for the Sommer parameter $r_0$ and the string tension $\sigma$
in units of $r_0$ from methods (i) and (ii), as explained in the text. Also
listed are the other fitparameters, the L\"uscher constant $\gamma$ and the
normalisation constant $V_0$.}
\label{tab:scale-fits}
\end{table}

\begin{figure}[t]
 \centering
\includegraphics[]{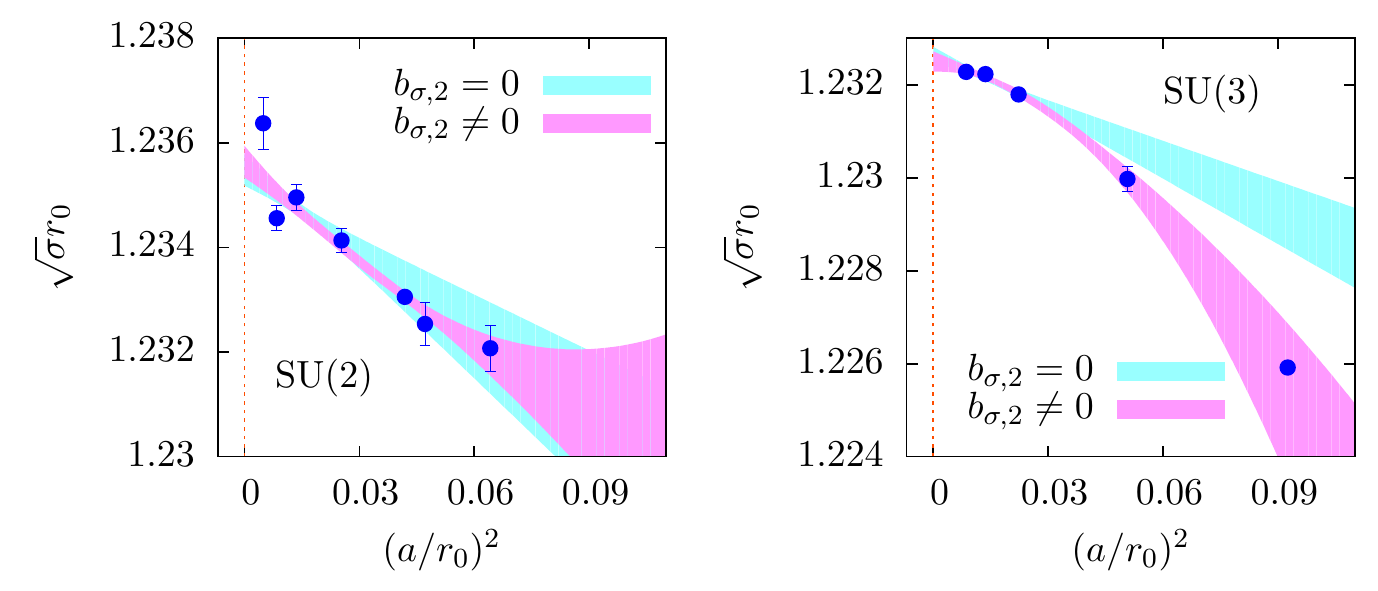}
 \caption{Continuum extrapolations of $\sqrt{\sigma_{(ii)}}r_0$ for SU(2) (left)
 and SU(3) gauge theory (right).}
 \label{fig:sig_conti}
\end{figure}

We want to obtain the asymptotic $R\to\infty$ behaviour of the potential in the
continuum. To this end we perform a continuum extrapolation for $\sqrt{\sigma}$
of the form,
\be
\label{eq:sig-conti}
\sqrt{\sigma} r_0 = \big( \sqrt{\sigma} r_0 \big)\cont + b_{\sigma,1}
\Big(\frac{a}{r_0}\Big)^2 + b_{\sigma,2} \Big(\frac{a}{r_0}\Big)^4 \,.
\ee
In practice, we perform two fits, one with $b_{\sigma,2}\neq0$ and another
one with $b_{\sigma,2}=0$. The continuum extrapolations for $\SU(2)$ and
$\SU(3)$ are displayed in figure~\ref{fig:sig_conti} and the results are tabulated
in table~\ref{tab:scale-conti}. For the fit with $b_{\sigma,2}=0$ and SU(3) we
could only include points with $(a/r_0)^2<0.03$ to obtain a reasonable
$\chi^2/$dof. We have applied a similar cut for the fit in the case of SU(2),
too. For SU(2) the three results at the smallest lattice spacings show
fluctuations, which for $\beta=16.0$ are bigger than $1\sigma$ with respect to
the fits, leading to a $\chi^2/$dof$>1$. The continuum results from the two fits
agree well in both cases. In the following we will use the result coming from
the analysis with $b_{\sigma,2}\neq0$.

\begin{table}
\centering
\begin{tabular}{c|cc|cc}
 \hline
 \hline
 & \multicolumn{2}{c|}{SU(2)} &
 \multicolumn{2}{c}{SU(3)} \\
 \hline
 \hline
 & quadratic & linear & quadratic & linear \\
 \hline
 $\big( \sqrt{\sigma} r_0 \big)\cont$ & 1.2356(3) & 1.2355(3) & 1.2325(3)
 & 1.2327(2) \\
 \hline
 \hline
\end{tabular}
\caption{Results from the continuum extrapolations with $b_{\sigma}=0$
(linear) and $b_{\sigma}\neq0$ (quadratic).}
\label{tab:scale-conti}
\end{table}

\subsection{The static potential}
\label{sec:res-pot}

\begin{figure}[t]
 \centering
\includegraphics[]{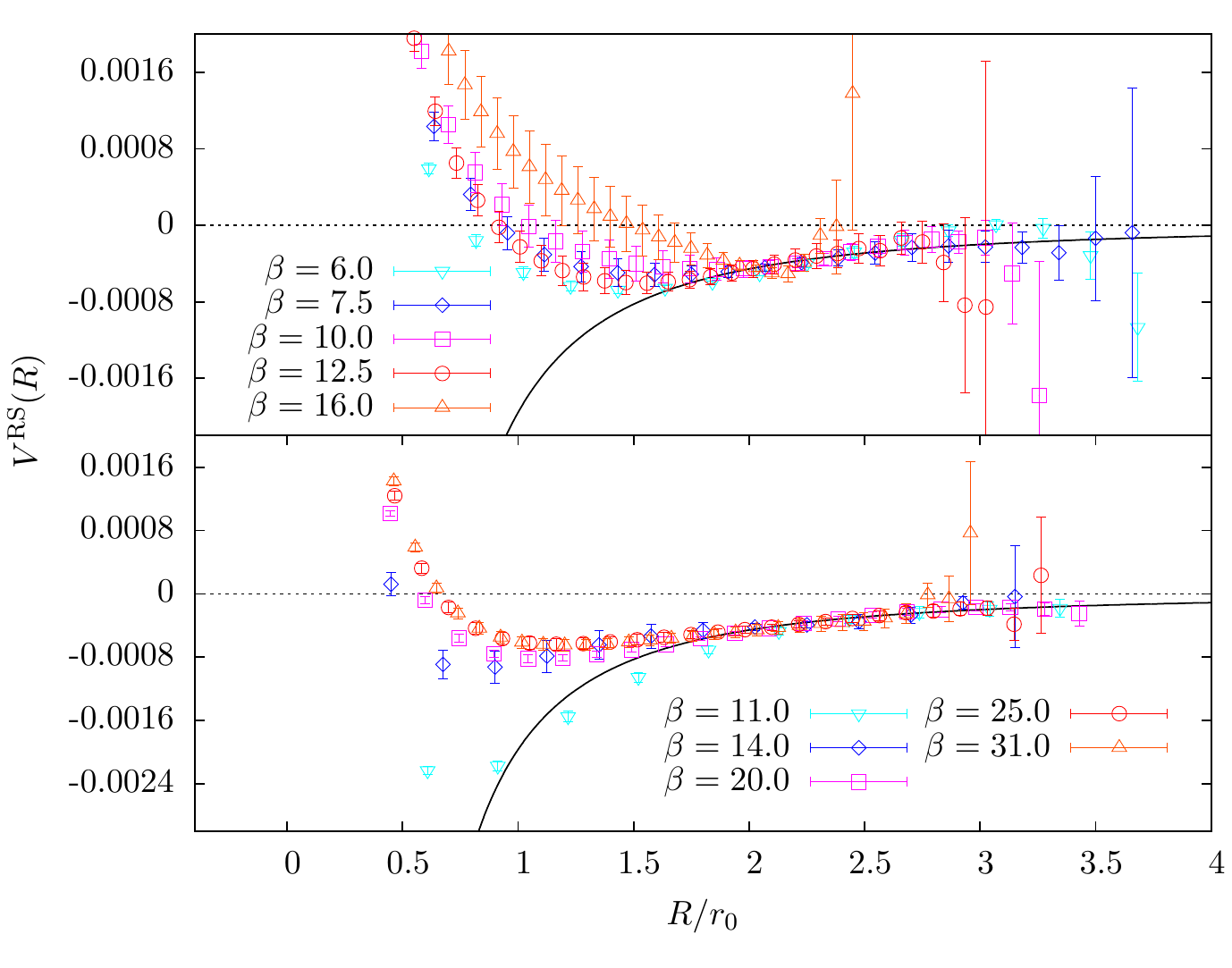}
 \caption{Results for the static potential (in its rescaled form -- see
eq.~\refc{eq:resc-pot}) for different lattice spacings versus $R$ in SU(2) (top)
and SU(3) (bottom) gauge theory. The black continuous line is the light cone
potential from eq.~\refc{eq:LC-spectrum} using the continuum extrapolated string
tension.}
 \label{fig:res_pot}
\end{figure}

In figure~\ref{fig:res_pot} we show the results for the static potential,
rescaled via
\be
\label{eq:resc-pot}
V^{\rm RS}(R) = \Big( \frac{V(R)-V_0}{\sqrt{\sigma}} - R\sqrt{\sigma} \Big)
\frac{R\sqrt{\sigma}}{\pi} + \frac{1}{24} \,.
\ee
In this rescaled form the leading order potential to $O(1/R)$ in the $1/R$
expansion is normalised to 0, so that small differences become visible. We have
rescaled the energies using the string tension for each individual value of
$\beta$. The solid lines in the plot correspond to the LC spectrum,
eq.~\refc{eq:LC-spectrum}, and are evaluated using the continuum extrapolated
string tension. For SU(2) gauge theory we have not plotted the potential for
the two largest lattice spacings for which the results look similar.

Corrections to the LC spectrum start at around $R/r_0\approx2$ (i.e. around 1~fm
in 4d physical units) and are positive, except for one case, namely
$\beta=11.0$ for SU(3). This means that the dominant corrections are not
expected to be due to the rigidity term from section~\ref{sec:beyondEST}. In
that case one would expect to obtain a negative correction for intermediate
values of $R/r_0$. However, this does not rule out the presence of the rigidity
term in general. It can still be a subleading correction for the lattice
spacings considered. The only exception is the SU(3) $\beta=11.0$ lattice
(with $a\approx0.15$~fm -- in physical units defined via $r_0\equiv0.5$~fm),
where the negative correction could be due to a dominant rigidity term. In all
of the cases we observe that the magnitude of the corrections becomes stronger
when we approach the continuum limit. In particular, they are larger for SU(2)
gauge theory than for the SU(3) case.

\subsection{Isolating the leading order correction terms}
\label{sec:LO-correction}

On top of the leading order behaviour (the LC spectrum from
eq.~\refc{eq:LC-spectrum}), the EST predicts corrections starting at $O(R^{-4})$.
We would now like to test whether this prediction is true. The
first step is to isolate the leading order correction to
eq.~\refc{eq:LC-spectrum}. If eq.~\refc{eq:LC-spectrum} is incorrect,
corrections will appear with an exponent $0<m<4$, if the EST predictions
are correct and $\bt\neq0$, we expect $m=4$, if corrections only appear at
higher orders we will obtain $m>4$. If eq.~\refc{eq:LC-spectrum} is correct to
all orders we should obtain $m\approx0$.

\begin{figure}[t]
 \centering
\includegraphics[]{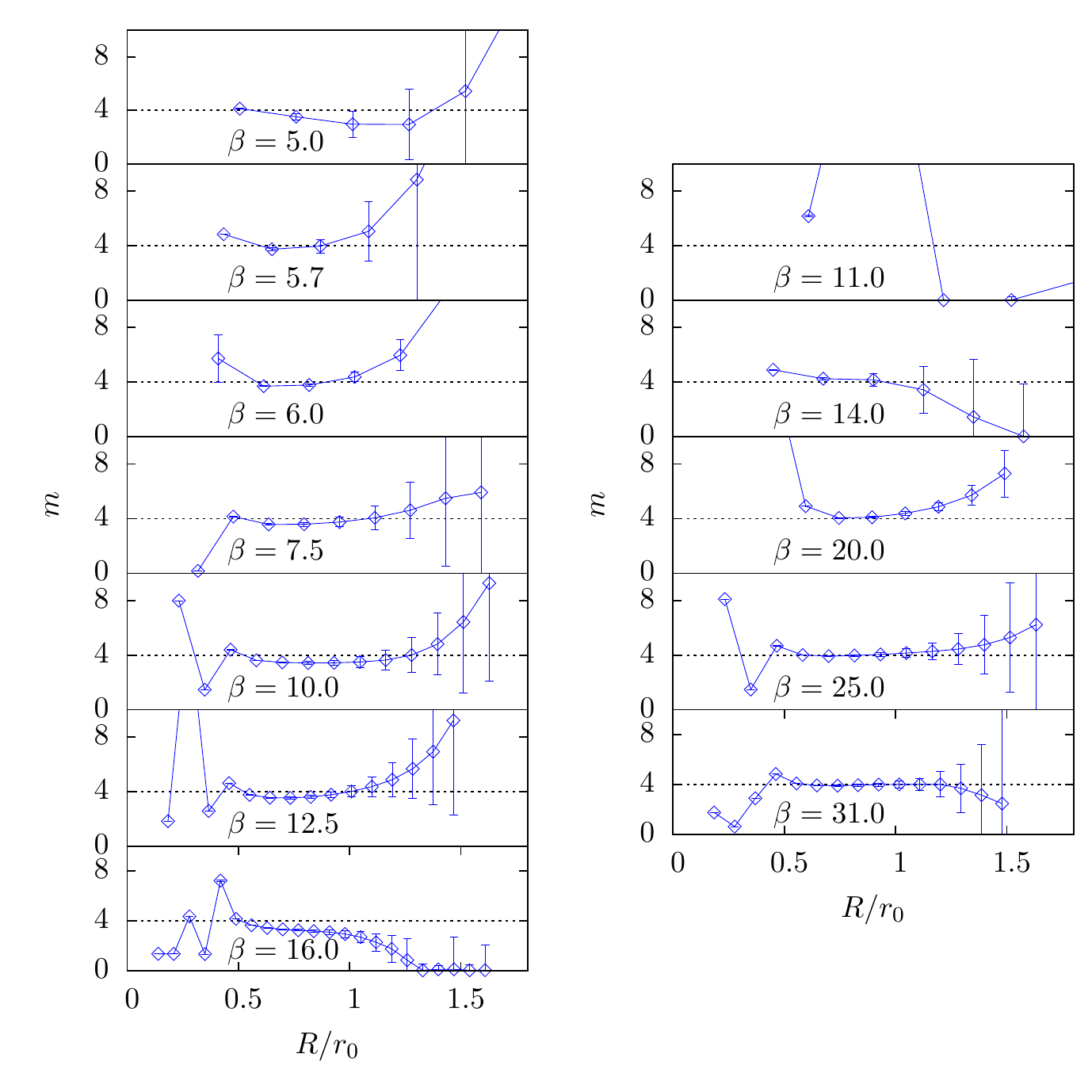}
 \caption{Results for the exponent $m$ plotted versus the minimal value of $R$
 included in the fit, $R_{\rm min}$, for SU(2) (left) and SU(3) (right) gauge
 theory. The horizontal line indicates the LO exponent of the EST.}
 \label{fig:m-exponent}
\end{figure}

To investigate the power of the leading order correction we fit the data to the
form
\be
\label{eq:lead-coeff-fit}
V(R) = E_0^{\rm LC}(R) + \frac{\eta}{\big(\sqrt{\sigma} R\big)^m} + V_0 \,,
\ee
where $E_0^{\rm LC}(R)$ is the potential obtained from the LC energy
levels~\refc{eq:LC-spectrum} and $\eta$ and $m$, together with $\sigma$ and
$V_0$, are fit parameters. The results for the exponent $m$ versus the minimal
value of $R$ included in the fit, $R_{\rm min}$, is shown in
figure~\ref{fig:m-exponent}. The results for the exponent typically show a
plateau in the region $0.5\lesssim R/r_0 \lesssim 1.0$ where $m\approx4$. The
only exception is, once more, $\beta=11.0$ for SU(3). For SU(2), we see that
the result for $m$ is typically a bit smaller than 4 (around 3.6).
This could indicate that we observe a mixing of two types of corrections where
the one of $O(R^{-4})$ is the dominant one. For $R_{\rm min}/r_0>1$ we cannot
resolve the correction terms reliably, so that $m$ is not determined
sufficiently.

\section{Analysis of boundary corrections without massive modes}
\label{sec:b2-ana1}

In this section we will now turn towards the extraction of the boundary
coefficient \btt. In this first part of the analysis we will neglect
contributions from massive modes (or the rigidity term) and extract the value of
\btt{} in this setup. We will see how \btt{} changes in the presence of these
terms in the next section.

\subsection{Extraction of the boundary coefficient}
\label{sec:b2-extract}

To extract \btt{} we use the groundstate energy from eq.~\refc{eq:est-spec}. To
check for the impact of higher order correction terms our general fit formula is
of the form
\be
\label{eq:boundary-fit}
V(R) = E^{\rm EST}_{0}(R) + \frac{\gamma^{(1)}_{0}}{\sqrt{\sigma^5} R^6} + 
\frac{\gamma^{(2)}_{0}}{\sigma^3 R^7} + V_0 \,,
\ee
where we have included appropriate powers of $\sigma$ multiplying the higher
order terms to keep the coefficients dimensionless. In practice, we perform
five different types of fits:
\begin{enumerate}
 \item[{\bf A}] we use the string tension and $V_0$ extracted from method
 (ii) in the determination of the string tension from
 section~\ref{sec:scale-setting} and use eq.~\refc{eq:boundary-fit} with \btt,
 $\gamma^{(1)}_{0}$ and $\gamma^{(2)}_{0}$ as free parameters;
 \item[{\bf B}] use $\sigma$, $V_0$  and \btt{} as free parameters and
 $\gamma^{(1)}_{0}=0$ and $\gamma^{(2)}_{0}=0$;
 \item[{\bf C}] use $\sigma$, $V_0$,  \btt{} and $\gamma^{(1)}_{0}$ as free
 parameters and $\gamma^{(2)}_{0}=0$;
 \item[{\bf D}] use $\sigma$, $V_0$, \btt{} and $\gamma^{(2)}_{0}$ as free
 parameters and $\gamma^{(1)}_{0}=0$;
 \item[{\bf E}] use $\sigma$, $V_0$, $\gamma^{(1)}_{0}$ and $\gamma^{(2)}_{0}=0$
 as free parameters and $\bt=0$.
\end{enumerate}

From the analysis in the previous section, we expect the correction terms to be
relevant starting with $R/r_0\approx1.2$. To test the region in which the data
is well described by the higher order terms we perform the fits for several
values of the lower cut in $R$ and check at which value for $R_{\rm min}$ we get
a good description, indicated by acceptable values for $\chi^2/$dof. For the
final result, i.e. for the $R_{\rm min}$ in the final fit, we pick the second
smallest distance for which the fit gave an acceptable $\chi^2/$dof$<1.5$. We use
the results with minimal $R$ value of $R_{\rm min}\pm1a$ to estimate the systematic
uncertainty of the extraction of the fit parameters for this particular fit.
For the fits including terms of $\Ord(R^{-6})$ or $\Ord(R^{-7})$ we would
expect that these work well down to even smaller values of $R_{\rm min}$ than
the fits including only the $R^{-4}$ term, since these fits include higher order
correction terms which should improve the agreement with the data.

\begin{table}
\small
\centering
\begin{tabular}{c|ll|l|ll|cc}
 \hline
 \hline
 Fit & \multicolumn{1}{c}{$\sqrt{\sigma}r_0$} & \multicolumn{1}{c|}{$aV_0$} &
\multicolumn{1}{c|}{$\bt\cdot10^{2}$} &
\multicolumn{1}{c}{$\gamma^{(1)}_{0}\cdot10^{3}$} &
\multicolumn{1}{c|}{$\gamma^{(2)}_{0}\cdot10^{3}$} & $\chi^2/$dof &
$R_{\rm
min}/r_0$ \\
 \hline
 \hline
 \multicolumn{2}{l}{{ }$\beta=5.0$} & & & & & & \\
 {\bf A} & \multicolumn{1}{c}{---} & \multicolumn{1}{c|}{---} & { }1.7(21)(23)
& { }7(21)(75) & -{ }6(17)(76) & 0.17 & 0.76 \\
 {\bf B} & 1.2138(2)(3) & 0.2151(1)(3) & -1.75({ }3)(22) &
\multicolumn{1}{c}{---} & \multicolumn{1}{c|}{---} & 0.22 & 0.76 \\
 {\bf C} & 1.2320(3)(3) & 0.2148(3)(4) & -2.29(32)(102) & -2({ }2)({ }7) &
\multicolumn{1}{c|}{---} & 0.12 & 0.76 \\
 {\bf D} & 1.2320(3)(3) & 0.2148(3)(3) & -2.15(24)(79) &
\multicolumn{1}{c}{---} & -1.3(7)(6) & 0.24 & 0.76 \\
 {\bf E} & 1.2321(3)(3) & 0.2148(5)(4) & \multicolumn{1}{c|}{---} & -67(73)(60) &
-64(71)(86) & 0.11 & 1.01 \\
 \hline
 \multicolumn{2}{l}{{ }$\beta=5.7$} & & & & & & \\
 {\bf A} & \multicolumn{1}{c}{---} & \multicolumn{1}{c|}{---} & { }1.6({ }7)(16)
& 60({ }4)(38) & -45({ }3)(31) & 0.01 & 0.87 \\
 {\bf B} & 1.2329(3)(1) & 0.2027(3)(1) & -1.96(21)(26) &
\multicolumn{1}{c}{---} & \multicolumn{1}{c|}{---} & 0.03 & 1.09 \\
 {\bf C} & 1.2329(4)(2) & 0.2027(4)(3) & -1.99(59)(114) & 0.3(25)(80) &
\multicolumn{1}{c|}{---} & 0.03 & 0.87 \\
 {\bf D} & 1.2329(4)(2) & 0.2027(4)(2) & -1.98(44)(80) &
\multicolumn{1}{c}{---} & -0.2(18)(68) & 0.03 & 0.87 \\
 {\bf E} & 1.2327(3)(3) & 0.2029(3)(3) & \multicolumn{1}{c|}{---} & 33(33)(12) &
-24(24)(12) & 0.01 & 0.87 \\
 \hline
 \multicolumn{2}{l}{{ }$\beta=6.0$} & & & & & & \\
 {\bf A} & \multicolumn{1}{c}{---} & \multicolumn{1}{c|}{---} & { }2.26(61)(140)
& 97({ }8)(47) & -80({ }5)(45) & 0.15 & 1.02 \\
 {\bf B} & 1.2333(1)(1) & 0.1973(1)(1) & -2.19(6)(12) &
\multicolumn{1}{c}{---} & \multicolumn{1}{c|}{---} & 0.25 & 1.02 \\
 {\bf C} & 1.2333(1)(1) & 0.1973(1)(1) & -2.27(15)(60) & -1.3(6)(41) &
\multicolumn{1}{c|}{---} & 0.28 & 0.82 \\
 {\bf D} & 1.2333(1)(1) & 0.1973(1)(1) & -2.20(12)(39) &
\multicolumn{1}{c}{---} & -0.9(4)(32) & 0.27 & 0.82 \\
 {\bf E} & 1.2332(2)(1) & 0.1975(2)(1) & \multicolumn{1}{c|}{---} & 43(44)(25) &
-35(36)(35) & 0.19 & 1.02 \\
 \hline
 \multicolumn{2}{l}{{ }$\beta=7.5$} & & & & & & \\
 {\bf A} & \multicolumn{1}{c}{---} & \multicolumn{1}{c|}{---} & -0.5(14)(14)
& 31(18)(39) & -22(12)(35) & 0.16 & 0.80 \\
 {\bf B} & 1.2342(2)(2) & 0.1738(1)(1) & -2.30(7)(17) &
\multicolumn{1}{c}{---} & \multicolumn{1}{c|}{---} & 0.05 & 0.95 \\
 {\bf C} & 1.2343(2)(1) & 0.1738(2)(1) & -2.69(20)(21) & -2.3(7)(9) &
\multicolumn{1}{c|}{---} & 0.03 & 0.80 \\
 {\bf D} & 1.2343(2)(1) & 0.1738(1)(1) & -2.54(14)(20) &
\multicolumn{1}{c}{---} & -1.6(5)(8) & 0.03 & 0.80 \\
 {\bf E} & 1.2342(2)(2) & 0.1739(2)(2) & \multicolumn{1}{c|}{---} & 52(53)(15) &
-43(43)(16) & 0.03 & 0.95 \\
 \hline
 \multicolumn{2}{l}{{ }$\beta=10.0$} & & & & & & \\
 {\bf A} & \multicolumn{1}{c}{---} & \multicolumn{1}{c|}{---} & -0.95(127)(117)
& 28(17)(33) & -21(11)(29) & 0.52 & 0.81 \\
 {\bf B} & 1.2350(1)(2) & 0.14486(5)(9) & -2.42(4)(17) &
\multicolumn{1}{c}{---} & \multicolumn{1}{c|}{---} & 0.28 & 0.93 \\
 {\bf C} & 1.2350(1)(3) & 0.14482(5)(13) & -2.78(6)(36) & 2.2(2)(12) &
\multicolumn{1}{c|}{---} & 0.19 & 0.70 \\
 {\bf D} & 1.2350(1)(3) & 0.14484(5)(14) & -2.58(5)(29) &
\multicolumn{1}{c}{---} & -1.3(1)(10) & 0.27 & 0.70 \\
 {\bf E} & 1.2349(2)(2) & 0.14491(7)(10) & \multicolumn{1}{c|}{---} & 55(55)(16)
& -45(45)(19) & 0.17 & 0.93 \\
 \hline
 \multicolumn{2}{l}{{ }$\beta=12.5$} & & & & & & \\
 {\bf A} & \multicolumn{1}{c}{---} & \multicolumn{1}{c|}{---} & 0.53(73)(96)
& 47(9)(25) & -33(6)(21) & 0.28 & 0.83 \\
 {\bf B} & 1.2347(1)(2) & 0.12467(4)(8) & -2.27(3)(12) &
\multicolumn{1}{c}{---} & \multicolumn{1}{c|}{---} & 0.33 & 0.83 \\
 {\bf C} & 1.2349(2)(1) & 0.12458(5)(5) & -2.75(9)(18) & -2.1(3)(7) &
\multicolumn{1}{c|}{---} & 0.06 & 0.73 \\
 {\bf D} & 1.2349(2)(1) & 0.12460(5)(5) & -2.58(7)(16) &
\multicolumn{1}{c}{---} & -1.3(2)(5) & 0.07 & 0.73 \\
 {\bf E} & 1.2346(2)(2) & 0.12476(5)(9) & \multicolumn{1}{c|}{---} & 39(39)(9)
& -28(28)(9) & 0.17 & 0.83 \\
 \hline
 \multicolumn{2}{l}{{ }$\beta=16.0$} & & & & & & \\
 {\bf A} & \multicolumn{1}{c}{---} & \multicolumn{1}{c|}{---} & -0.2(8)(110)
& 60(9)(36) & -50(5)(33) & 1.20 & 0.91 \\
 {\bf B} & 1.2357(1)(2) & 0.10281(3)(5) & -2.56(3)(14) &
\multicolumn{1}{c}{---} & \multicolumn{1}{c|}{---} & 0.95 & 0.91 \\
 {\bf C} & 1.2358(1)(3) & 0.10277(3)(7) & -3.01(5)(28) & -2.6(2)(10) &
\multicolumn{1}{c|}{---} & 0.76 & 0.70 \\
 {\bf D} & 1.2358(1)(2) & 0.10275(3)(5) & -2.97(6)(22) &
\multicolumn{1}{c}{---} & -2.2(2)(10) & 0.54 & 0.77 \\
 {\bf E} & 1.2356(2)(2) & 0.10285(4)(7) & \multicolumn{1}{c|}{---} & 55(55)(13)
& -44(44)(14) & 0.63 & 0.91 \\
 \hline
 \hline
\end{tabular}
\caption{Results of the fits for the extraction of \btt{} for $\SU(2)$ gauge
theory.}
\label{tab:b2-fits-su2}
\end{table}

\begin{table}
\small
\centering
\begin{tabular}{c|ll|l|ll|cc}
 \hline
 \hline
 Fit & \multicolumn{1}{c}{$\sqrt{\sigma}r_0$} & \multicolumn{1}{c|}{$aV_0$} &
\multicolumn{1}{c|}{$\bt\cdot10^{2}$} &
\multicolumn{1}{c}{$\gamma^{(1)}_{0}\cdot10^{3}$} &
\multicolumn{1}{c|}{$\gamma^{(2)}_{0}\cdot10^{3}$} & $\chi^2/$dof &
$R_{\rm min}/r_0$ \\
 \hline
 \hline
 \multicolumn{2}{l}{{ }$\beta=11.0$} & & & & & & \\
 {\bf A} & \multicolumn{1}{c}{---} & \multicolumn{1}{c|}{---} & { }1.6({ }5)(19)
& { }14(8)(92) & -{ }3(6)(100) & 0.07 & 0.91 \\
 {\bf B} & 1.2259(1)(1) & 0.2498(1)(2) & { }1.0({ }2)({ }9) &
\multicolumn{1}{c}{---} & \multicolumn{1}{c|}{---} & 0.02 & 1.52 \\
 {\bf C} & 1.2260(1)(2) & 0.2497(1)(4) & { }1.3({ }2)(13) & { }10(6)(12) &
\multicolumn{1}{c|}{---} & 0.05 & 0.91 \\
 {\bf D} & 1.2260(1)(2) & 0.2497(1)(4) & { }0.8(10)(11) &
\multicolumn{1}{c}{---} & { }{ }8(1)({ }12) & 0.07 & 0.91 \\
 {\bf E} & 1.2260(1)(2) & 0.2496(1)(3) & \multicolumn{1}{c|}{---} & -17(2)(46) &
{ }22(2)({ }64) & 0.12 & 0.91 \\
 \hline
 \multicolumn{2}{l}{{ }$\beta=14.0$} & & & & & & \\
 {\bf A} & \multicolumn{1}{c}{---} & \multicolumn{1}{c|}{---} & { }0.81(15)(43)
& { }{ }6(10)(11) & -{ }3(6)({ }10) & 0.05 & 0.68 \\
 {\bf B} & 1.2300(2)(1) & 0.2239(1)(2) & -1.37({ }4)(10) &
\multicolumn{1}{c}{---} & \multicolumn{1}{c|}{---} & 0.02 & 0.90 \\
 {\bf C} & 1.2300(2)(2) & 0.2239(2)(2) & -1.24({ }9)(30) & 0.7(3)({ }5) &
\multicolumn{1}{c|}{---} & 0.01 & 0.68 \\
 {\bf D} & 1.2300(2)(1) & 0.2239(2)(1) & -1.29({ }7)(10) &
\multicolumn{1}{c}{---} & 0.4(2)({ }{ }3) & 0.01 & 0.68 \\
 {\bf E} & 1.2299(2)(1) & 0.2240(2)(2) & \multicolumn{1}{c|}{---} & {
}23(5)(18) & -17(5)({ }23) & 0.03 & 0.90 \\
 \hline
 \multicolumn{2}{l}{{ }$\beta=20.0$} & & & & & & \\
 {\bf A} & \multicolumn{1}{c}{---} & \multicolumn{1}{c|}{---} & -0.44(26)(63)
& { }17(3)(18) & -11(2)({ }15) & 0.11 & 0.75 \\
 {\bf B} & 1.2319(2)(0) & 0.18166(4)(1) & -1.71({ }2)({ }1) &
\multicolumn{1}{c}{---} & \multicolumn{1}{c|}{---} & 0.003 & 0.89 \\
 {\bf C} & 1.2319(2)(0) & 0.18166(5)(1) & -1.73({ }6)({ }8) & -0.1(2)({ }2) &
\multicolumn{1}{c|}{---} & 0.004 & 0.75 \\
 {\bf D} & 1.2319(2)(0) & 0.18166(5)(1) & -1.72({ }4)({ }5) &
\multicolumn{1}{c}{---} & -0.1(1)({ }{ }1) & 0.004 & 0.75 \\
 {\bf E} & 1.2318(2)(1) & 0.18178(5)(7) & \multicolumn{1}{c|}{---} & {
}30(2)({ }8) & -22(2)({ }{ }9) & 0.02 & 0.89 \\
 \hline
 \multicolumn{2}{l}{{ }$\beta=25.0$} & & & & & & \\
 {\bf A} & \multicolumn{1}{c}{---} & \multicolumn{1}{c|}{---} & 0.2({ }{
}4)(110) & { }41(5)(48) & -32(3)({ }53) & 0.04 & 0.93 \\
 {\bf B} & 1.2319(1)(0) & 0.15696(3)(2) & -1.78({ }2)({ }4) &
\multicolumn{1}{c}{---} & \multicolumn{1}{c|}{---} & 0.004 & 0.93 \\
 {\bf C} & 1.2319(1)(1) & 0.15695(3)(10) & -1.83({ }4)(32) & -0.3(2)({ }5) &
\multicolumn{1}{c|}{---} & 0.004 & 0.70 \\
 {\bf D} & 1.2319(1)(1) & 0.15695(3)(8) & -1.81({ }3)(19) &
\multicolumn{1}{c}{---} & -0.2(1)({ }{ }5) & 0.004 & 0.70 \\
 {\bf E} & 1.2318(2)(1) & 0.15704(4)(7) & \multicolumn{1}{c|}{---} & {
}36(3)(17) & -28(3)({ }23) & 0.03 & 0.93 \\
 \hline
 \multicolumn{2}{l}{{ }$\beta=31.0$} & & & & & & \\
 {\bf A} & \multicolumn{1}{c}{---} & \multicolumn{1}{c|}{---} & 0.0({ }{
}4)({ }6) & { }28(4)(15) & -19(3)({ }13) & 0.55 & 0.83 \\
 {\bf B} & 1.2324(1)(1) & 0.13542(2)(2) & -1.77({ }1)({ }2) &
\multicolumn{1}{c}{---} & \multicolumn{1}{c|}{---} & 0.34 & 0.74 \\
 {\bf C} & 1.2324(1)(2) & 0.13540(2)(6) & -1.84({ }2)(15) & -0.23(5)(41) &
\multicolumn{1}{c|}{---} & 0.24 & 0.65 \\
 {\bf D} & 1.2324(1)(1) & 0.13541(2)(5) & -1.81({ }8)(10) &
\multicolumn{1}{c}{---} & -0.12(3)({ }21) & 0.25 & 0.65 \\
 {\bf E} & 1.2322(1)(1) & 0.13552(2)(5) & \multicolumn{1}{c|}{---} & {
}27(28)({ }5) & -19(19)({ }6) & 0.46 & 0.83 \\
 \hline
 \hline
\end{tabular}
\caption{Results of the fits for the extraction of \btt{} for $\SU(3)$ gauge
theory.}
\label{tab:b2-fits-su3}
\end{table}

The results of the fits are listed in table~\ref{tab:b2-fits-su2} for $\SU(2)$
gauge theory and in table~\ref{tab:b2-fits-su3} for the $\SU(3)$ case. Let us
discuss the individual fits before moving on with the analysis. In general, the
fits lead to small values of $\chi^2/$dof (since we picked the second possible
value for $R_{\rm min}$), so that it is difficult to make
statements about the agreement of the EST predictions with the data based on
these numbers. Accordingly, our main argument to judge the agreement will be
based on the values of $R_{\rm min}$ which can be used in the fits. At coarse
lattice spacings all fits work equally well, leading to similar of $R_{\rm
min}$. When going to smaller lattice spacings, however, the picture changes
slightly. In these cases the worst fits are typically fit {\bf B} (which is not
surprising since higher order terms have been neglected) and fits {\bf A} and
{\bf E}. For fit {\bf A} this might be due to the fact that the fit does not
allow $\sigma$ and $V_0$ to change, which, apparently, is to restrictive, even
though both do not change significantly for the other fits. For fit {\bf E}
$R_{\rm min}$ needs to be larger even though higher order terms are included in
the fit, indicating less agreement with the data. This implies that the
scenario with $\bt=0$ is disfavoured, in agreement with the analysis from
section~\ref{sec:LO-correction}. For the following analysis we will thus include
the results from fits {\bf C} and {\bf D} together with the one from fit
{\bf B}, for which the larger value of $R_{\rm min}$ has been expected. We note,
however, that we cannot fully exclude the possibility that the
coefficient of the $R^{-4}$ term is 0 and that the result from
section~\ref{sec:LO-correction} is due to a fine-tuned combination of higher
order terms.

\begin{table}
\small
\centering
\begin{tabular}{cc|cc}
 \hline
 \hline
 \multicolumn{2}{c|}{$\SU(2)$} & \multicolumn{2}{c}{$\SU(3)$} \\
 $\beta$ & \btt & $\beta$ & \btt \\
 \hline
 \hline
 5.0 & -0.0179({ }5)(50)(23) & 11.0 & 0.0100(15)(34)(95) \\
 5.7 & -0.0196(25)({ }3)({ }2) & 14.0 & -0.0133({ }6)(10)({ }2) \\
 6.0 & -0.0211({ }7)(17)(11) & 20.0 & -0.0171({ }3)({ }7)({ }2) \\
 7.5 & -0.0244(11)(25)(16) & 25.0 & -0.0175({ }8)({ }7)({ }2) \\
 10.0 & -0.0251({ }5)(27)(22) & 31.0 & -0.0178({ }6)({ }7)({ }3) \\
 12.5 & -0.0245({ }5)(31)(13) & & \\
 16.0 & -0.0273({ }4)(29)(17) & & \\
 \hline
 \hline
\end{tabular}
\caption{Final results for \btt{} for the individual lattice spacings. The
first error is the statistical uncertainty, the second the systematic one due to
the unknown correction terms, estimated by calculating the maximal deviations of
the results for \btt{} from fits {\bf B} to {\bf C}, and the third is the
systematic one associated with the choice for $R_{\rm min}$.}
\label{tab:b2-results}
\end{table}

We determine the associated result for \btt{} on a single lattice spacing
using the average over the results from fits {\bf B} to {\bf D}, weighted with
the associated uncertainties. To determine the systematic error due to the
particular choice for $R_{\rm min}$ we repeat the same procedure with $R_{\rm
min}\pm 1a$ and take the maximal deviation as an estimate. The results are
listed in table~\ref{tab:b2-results}. In the next section we discuss the
associated continuum extrapolation. It is interesting to note that in $\SU(3)$
gauge theory the systematic errors are typically an order of magnitude smaller
than in the $\SU(2)$ gauge theory. A possible reason could be, that the
agreement between the effective string theory predictions and the energy levels
becomes better with increasing $N$. This appears natural in the light of
possible corrections to the EST discussed in section~\ref{sec:est-setup}, which
are expected to be further suppressed with increasing $N$.

\begin{figure}[t]
 \centering
\includegraphics[]{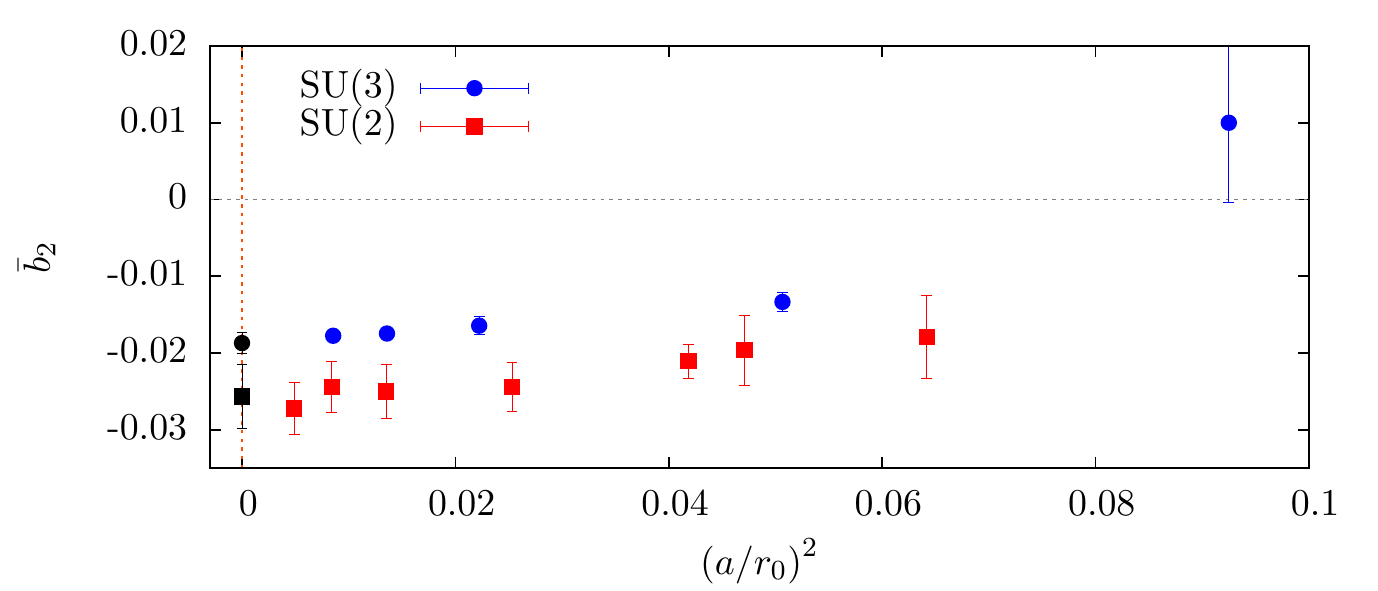}
 \caption{Results for \btt{} from table~\ref{tab:b2-results} versus the
 squared lattice spacing in units of $r_0$. Also shown are the continuum
 results from eq.~\refc{eq:b2-conti-res}.}
 \label{fig:br_vs_r0}
\end{figure}

The results for \btt{} are plotted versus $a^2$ in figure~\ref{fig:br_vs_r0}.
The plot displays the rather smooth behaviour towards the continuum ($a=0$).
The exception is the data point at $\beta=11.0$ for gauge group $\SU(3)$,
which shows a rather strong upwards trend. We have
excluded this data point from the following analysis, since it appears to
lie outside of the scaling region.

\subsection{Continuum limit of boundary corrections}
\label{sec:b2-conti}

Up to now the comparison with the EST has been done at finite lattice spacing.
To extract the final continuum results we have to perform the continuum
extrapolation. To this end we parameterise the boundary coefficient \btt{} as
\be
\label{eq:b2-conti}
\bt = \big( \bt \big)\cont + b_{\bt,1} \Big(\frac{a}{r_0}\Big)^2 + b_{\bt,2}
\Big(\frac{a}{r_0}\Big)^4 \,.
\ee
Using this parameterisation we now perform three different fits:
\begin{enumerate}
 \item[(1)] a fit including all terms in eq.~\refc{eq:b2-conti} and all
 lattice spacings (except for $\beta=11.0$ for $\SU(3)$);
 \item[(2)] fit (1) but with $b_{\bt,2}=0$;
 \item[(3)] a fit with $b_{\bt,2}=0$, including only the $\beta$-values
 7.5, 10.0, 12.5 and 16.0 for $\SU(2)$ gauge theory and 20.0, 25.0
 and 31.0 for $\SU(3)$.
\end{enumerate}
As discussed above, we have excluded the data set with $\beta=11.0$. In
all cases we have included systematic errors due to the functional form and
the particular value for $R_{\rm min}$ used for the extraction of \btt{} by
performing these fits for the results from fits {\bf B} to {\bf D}
and the fits with a minimal $R$ value of $R_{\rm min}\pm1a$ individually.

\begin{figure}[t]
 \centering
\includegraphics[]{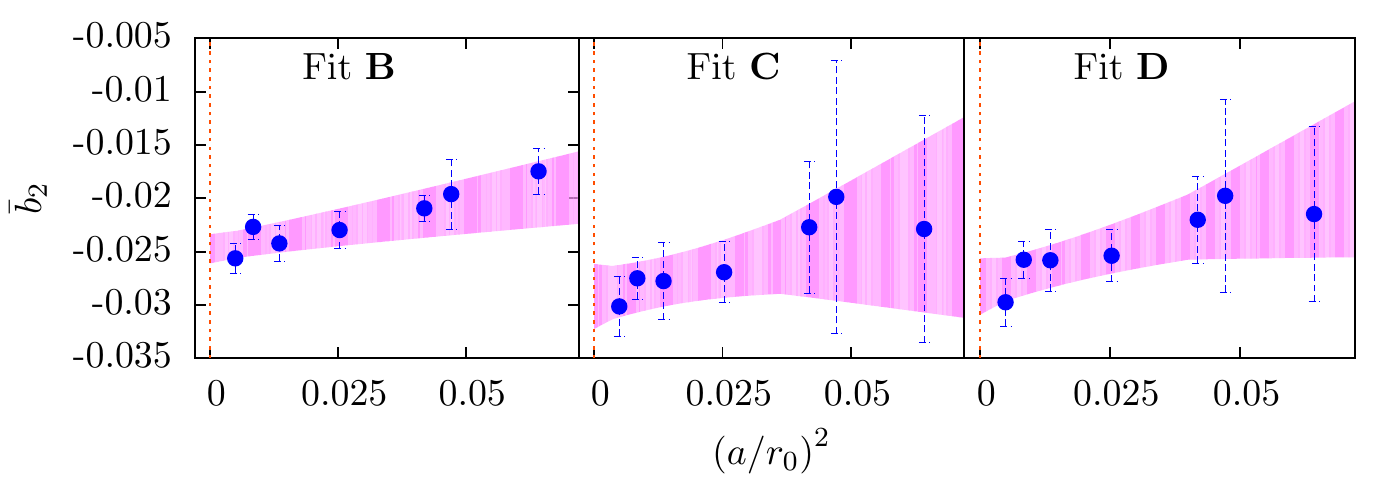} \\
\includegraphics[]{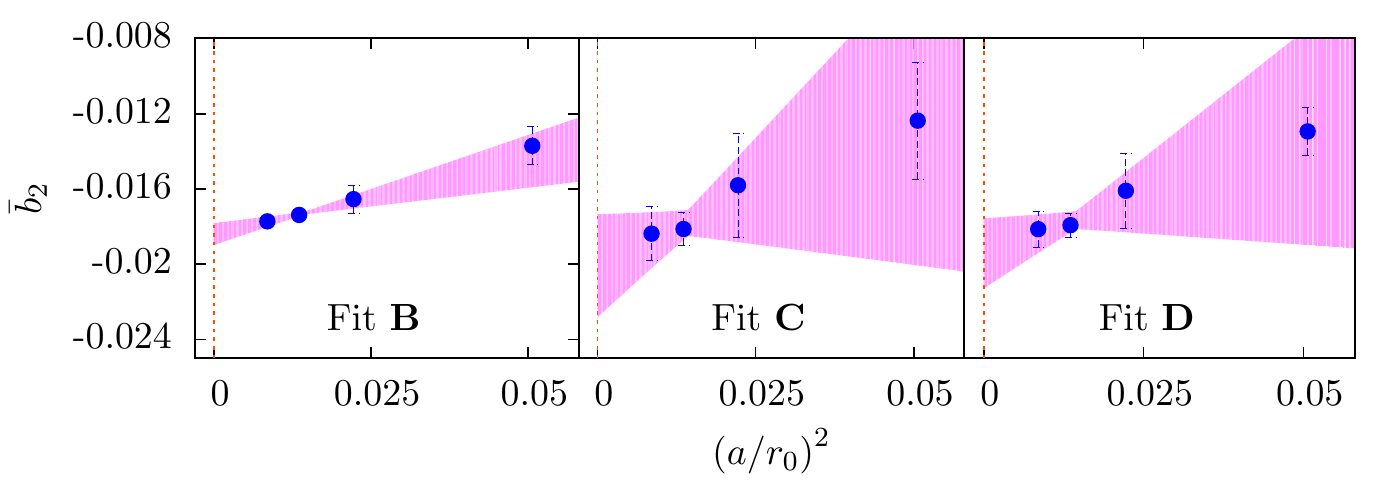} \\
 \caption{Results for the linear continuum extrapolation, fit (2), for the
results for \btt{} obtained from fits {\bf B}, {\bf C} and {\bf D} (from left to
right) for $\SU(2)$ (top) and $\SU(3)$ (bottom) gauge theory.}
 \label{fig:br_vs_r0_fit}
\end{figure}

To determine the final result, we have averaged the results weighted with the
individual uncertainties of the extrapolations for fits {\bf B} to {\bf C} and
estimated the systematic uncertainty due to the choice for $R_{\rm min}$ by
doing the same for $R_{\rm min}\pm1a$. The procedure is the same as the one used
for the averaging at the individual lattice spacings. In this way we obtain a
final result for the three different fits (1) to (3). The continuum results for
\btt{} for the different fits are given in table~\ref{tab:b2_conti_fitpar}. As
can be seen from figure~\ref{fig:br_vs_r0}, the data is consistent with a
straight line, which is also indicated by good values for $\chi^2/$dof, below
but close to 1, for fit (2), even though they still might show a slight
curvature. As our final result we thus use the result from fit (2), including
the spread of the results from fits (1) and (3) as a systematic uncertainty
associated with the continuum extrapolation. The associated curves from fit (2)
with the main value for $R_{\rm min}$ are shown in
figure~\ref{fig:br_vs_r0_fit}.

\begin{table}
\small
\centering
\begin{tabular}{cc|ccc}
 \hline
 \hline
 $N$ & Fit & (1) & (2) & (3) \\
 \hline
 2 & $\big( \bt \big)\cont$ & -0.0254(5)(42)(27) & -0.0257(3)(38)(17) &
 -0.0256(5)(37)(16) \\
 3 & $\big( \bt \big)\cont$ & -0.0185(3)(14)(10) & -0.0187(2)(13)({ }4) &
 -0.0186(2)(15)({ }8) \\
 \hline
 \hline
\end{tabular}
\caption{Results for $\big( \bt \big)\cont$ from fits (1) to (3) (see text).
The first error is the statistical uncertainty, the second one the one
associated with the unknown higher order correction terms in the potential and
the third one the one due to the choice for $R_{\rm min}$.}
\label{tab:b2_conti_fitpar}
\end{table}

The final continuum results for \btt{}, given in eq.~\refc{eq:b2-conti-res}
in section~\ref{sec:results}, are already shown in figure~\ref{fig:br_vs_r0}.
The figure indicates that the results for \btt{} in $\SU(3)$
gauge theory are significantly larger than the results for $\SU(2)$,
indicating the non-universality of the boundary corrections in the EST.
In particular, this difference remains in the continuum limit.
From the tendency between $\SU(2)$ and $\SU(3)$, one might think that \btt{}
tends to zero for $N\to\infty$. This will be investigated further in the
next publication of the series.

\section{Analysis of boundary corrections with massive modes}
\label{sec:rigidity}

Up to now we have ignored possible contributions from massive modes.
Our aim in this section is to see how the
presence of these modes changes the result for \btt{} from the previous
section.

\subsection{Testing the consistency with the potential}
\label{sec:m-test}

To include the massive modes in the analysis we include the terms from
eqs.~\refc{eq:pot-rigid} and~\refc{eq:pot-rigid-cor} in the fit function
eq.~\refc{eq:boundary-fit},
\be
\label{eq:mass-fit}
V(R) = E_{0}\eff(R) - \frac{m}{2\pi} \sum_{k=1}^{\infty} \frac{K_1(2kmR)}{k}
- \frac{(d-2)(d-10)\pi^2}{3840 m \sigma R^4} +
\frac{\gamma^{(1)}_{0}}{\sqrt{\sigma^5} R^6}
+ \frac{\gamma^{(2)}_{0}}{\sigma^3 R^7} + V_0 \,.
\ee
The main difficulty when fitting to eq.~\refc{eq:mass-fit} is the presence
of the infinite sum over the modified Bessel functions of the second kind.
In practice, however, the sum is always completely dominated by the first
few terms, since $K_1(nc)$ decays exponentially with increasing $n$ for
any positive real number $c$. In our fits we have used the first 100 terms
and checked explicitly the correction to this approximation by calculating
the corrections due to the next 10 terms in the sum. For all combinations
of $m$ and $R$ that appeared in our analysis the correction has been
suppressed by 100 orders of magnitude, at least.

Using eq.~\refc{eq:mass-fit} we have performed the four different types
of fits:
\begin{enumerate}
 \item[{\bf F}] use $\sigma$, $V_0$, \btt{} and $m$ as free parameters and
 $\gamma^{(1)}_{0}=\gamma^{(2)}_{0}=0$;
 \item[{\bf G}] use $\sigma$, $V_0$, \btt{}, $m$ and $\gamma^{(1)}_{0}$ as free
 parameters and  $\gamma^{(2)}_{0}=0$;
 \item[{\bf H}] use $\sigma$, $V_0$, \btt{}, $m$ and $\gamma^{(2)}_{0}$ as free
 parameters and  $\gamma^{(1)}_{0}=0$;
 \item[{\bf J}] use $\sigma$, $V_0$ and $m$ as free parameters and
 $\gamma^{(1)}_{0}=\gamma^{(2)}_{0}=\bt=0$;
\end{enumerate}
The last fit constitutes a check whether the presence of the $R^{-4}$
term due to the massive mode (i.e. the term from
eq.~\refc{eq:pot-rigid-cor}) is already sufficient to describe the $R^{-4}$
correction found in section~\ref{sec:LO-correction}.

\begin{table}
\small
\centering
\begin{tabular}{c|ll|l|l|l|cc}
 \hline
 \hline
 Fit & \multicolumn{1}{c}{$\sqrt{\sigma}r_0$} & \multicolumn{1}{c|}{$aV_0$} &
\multicolumn{1}{c|}{$\bt\cdot10^{2}$} & \multicolumn{1}{c|}{$r_0m$} &
\multicolumn{1}{c}{$\gamma^{(1)/(2)}_{0}\cdot10^{3}$} & $\chi^2/$dof &
$R_{\rm min}/r_0$ \\
 \hline
 \hline
 \multicolumn{2}{l}{{ }$\beta=5.0$} & & & & & & \\
 {\bf F} & 1.2320(2)(1) & 0.2148(2)(2) & -0.8({ }1)({ }4) & 2.75({ }6)(49) &
 \multicolumn{1}{c|}{---} & 0.12 & 0.76 \\
 {\bf G} & 1.2322(4)(2) & 0.2145(5)(3) & -2.0({ }3)({ }1) & 2.01(70)(45) &
 -6.3(12)(56) & 0.11 & 0.76 \\
 {\bf H} & 1.2322(3)(2) & 0.2145(4)(3) & -1.7(11)({ }8) & 2.06(44)(50) &
 -3.6(52)(35) & 0.10 & 0.76 \\
 {\bf J} & 1.2320(3)(2) & 0.2149(3)(3) & \multicolumn{1}{c|}{---} &
 2.63({ }7)(22) & \multicolumn{1}{c|}{---} & 0.19 & 1.52 \\
 \hline
 \multicolumn{2}{l}{{ }$\beta=5.7$} & & & & & & \\
 {\bf F} & 1.2331(10)(2) & 0.2024(12)(3) & -1.3({ }8)({ }6) & 2.1(149)(607) &
 \multicolumn{1}{c|}{---} & 0.06 & 0.87 \\
 {\bf G} & 1.2328(8)(5) & 0.2026(10)(7) & -7.4(321)(73) & 1.2(150)(40) &
 -45(220)(81) & 0.02 & 0.87 \\
 {\bf H} & 1.2329(5)(3) & 0.2027(7)(3) & -1.2(17)({ }7) & 5.5(69)(24) &
 -0.8(71)(19) & 0.03 & 0.87 \\
 {\bf J} & 1.2327(2)(4) & 0.2029(2)(4) & \multicolumn{1}{c|}{---} &
 2.93(19)(89) & \multicolumn{1}{c|}{---} & 0.06 & 1.30 \\
 \hline
 \multicolumn{2}{l}{{ }$\beta=6.0$} & & & & & & \\
 {\bf F} & 1.2333(1)(1) & 0.1973(1)(1) & -1.0({ }1)(10) & 3.5({ }5)(238) &
 \multicolumn{1}{c|}{---} & 0.29 & 0.82 \\
 {\bf G} & 1.2333(2)(2) & 0.1972(1)(3) & -6.7({ }6)(69) & 1.3({ }1)(27) &
 -37(42)(710) & 0.33 & 0.82 \\
 {\bf H} & 1.2333(2)(2) & 0.1972(2)(3) & -4.9({ }5)(52) & 1.4({ }1)(25) &
 -21({ }3)(57) & 0.34 & 0.82 \\
 {\bf J} & 1.2332(1)(2) & 0.1975(1)(2) & \multicolumn{1}{c|}{---} &
 2.88(33)(454) & \multicolumn{1}{c|}{---} & 0.27 & 1.43 \\
 \hline
 \multicolumn{2}{l}{{ }$\beta=7.5$} & & & & & & \\
 {\bf F} & 1.2343(2)(3) & 0.1738(1)(2) & -1.15({ }5)(14) & 2.9({ }2)({ }6) &
 \multicolumn{1}{c|}{---} & 0.08 & 0.80 \\
 {\bf G} & 1.2345(3)(2) & 0.1736(4)(2) & -5.1(63)(38) & 1.5(37)(18) &
 -24(37)(30) & 0.06 & 0.80 \\
 {\bf H} & 1.2345(5)(2) & 0.1736(5)(3) & -3.7(53)(26) & 1.6(68)(15) &
 -12(25)(22) & 0.06 & 0.80 \\
 {\bf J} & 1.2341(2)(2) & 0.1740(1)(2) & \multicolumn{1}{c|}{---} &
 2.70({ }8)(15) & \multicolumn{1}{c|}{---} & 0.11 & 1.43 \\
 \hline
 \multicolumn{2}{l}{{ }$\beta=10.0$} & & & & & & \\
 {\bf F} & 1.2351(1)(2) & 0.14475(6)(8) & -1.47(11)(18) & 2.6({ }2)({ }3) &
 \multicolumn{1}{c|}{---} & 0.10 & 0.93 \\
 {\bf G} & 1.2352(2)(2) & 0.14472(2)(2) & -1.74(42)(23) & 2.7({ }6)(10) &
 -3({ }2)(13) & 0.06 & 0.70 \\
 {\bf H} & 1.2352(2)(2) & 0.1447{ }(2)(2) & -1.67(33)(29) & 2.6({ }4)({ }3) &
 -2({ }2)({ }2) & 0.07 & 0.70 \\
 {\bf J} & 1.2348(1)(2) & 0.14497(5)(9) & \multicolumn{1}{c|}{---} &
 2.75({ }3)(11) & \multicolumn{1}{c|}{---} & 0.50 & 1.40 \\
 \hline
 \multicolumn{2}{l}{{ }$\beta=12.5$} & & & & & & \\
 {\bf F} & 1.2349(2)(2) & 0.12459(5)(6) & -1.27({ }5)({ }9) & 3.0({ }3)({ }4) &
 \multicolumn{1}{c|}{---} & 0.11 & 0.83 \\
 {\bf G} & 1.2349(2)(2) & 0.12456(7)(5) & -1.49(11)(13) & 3.2({ }4)({ }3) &
 -2.0({ }6)({ }8) & 0.08 & 0.64 \\
 {\bf H} & 1.2349(2)(2) & 0.12456(7)(8) & -1.40(11)(16) & 3.0({ }4)({ }1) &
 -1.2({ }4)({ }7) & 0.09 & 0.64 \\
 {\bf J} & 1.2345(2)(2) & 0.12478(4)(7) & \multicolumn{1}{c|}{---} &
 2.88({ }3)({ }9) & \multicolumn{1}{c|}{---} & 0.42 & 1.28 \\
 \hline
 \multicolumn{2}{l}{{ }$\beta=16.0$} & & & & & & \\
 {\bf F} & 1.2359(1)(2) & 0.10273(3)(5) & -1.62({ }4)(18) & 2.7({ }1)({ }2) &
 \multicolumn{1}{c|}{---} & 0.60 & 0.91 \\
 {\bf G} & 1.2361(1)(2) & 0.10265(3)(7) & -2.34({ }9)(49) & 2.4({ }1)({ }3) &
 -6.0({ }5)(28) & 0.35 & 0.70 \\
 {\bf H} & 1.2360(1)(2) & 0.10269(3)(7) & -1.93({ }6)(35) & 2.5({ }1)({ }3) &
 -3.0({ }2)(17) & 0.47 & 0.70 \\
 {\bf J} & 1.2356(2)(2) & 0.10285(4)(4) & \multicolumn{1}{c|}{---} &
 2.74({ }1)({ }6) & \multicolumn{1}{c|}{---} & 0.88 & 1.40 \\
 \hline
 \hline
\end{tabular}
\caption{Results of the fits for the extraction of \btt{} and $m$ for
$\SU(2)$ gauge theory.}
\label{tab:b2m-fits-su2}
\end{table}

\begin{table}
\small
\centering
\begin{tabular}{c|ll|l|l|l|cc}
 \hline
 \hline
 Fit & \multicolumn{1}{c}{$\sqrt{\sigma}r_0$} & \multicolumn{1}{c|}{$aV_0$} &
\multicolumn{1}{c|}{$\bt\cdot10^{2}$} & \multicolumn{1}{c|}{$r_0m$} &
\multicolumn{1}{c}{$\gamma^{(1)/(2)}_{0}\cdot10^{3}$} & $\chi^2/$dof &
$R_{\rm min}/r_0$ \\
 \hline
 \hline
 \multicolumn{2}{l}{{ }$\beta=11.0$} & & & & & & \\
 {\bf F} & 1.2261(1)(3) & 0.2493(1)(5) & -1.11({ }3)(15) & 1.46({ }2)(32) &
 \multicolumn{1}{c|}{---} & 0.20 & 0.91 \\
 {\bf G} & 1.2260(1)(1) & 0.2495(2)(2) & -6.2{ }({ }4)(62) & 1.11({ }3)(25) &
 -44({ }4)(82) & 0.01 & 0.91 \\
 {\bf H} & 1.2260(1)(1) & 0.2495(2)(2) & -4.6{ }({ }3)(47) & 1.15({ }3)(24) &
 -30({ }3)(76) & 0.01 & 0.91 \\
 {\bf J} & 1.2260(1)(2) & 0.2495(1)(3) & \multicolumn{1}{c|}{---} &
 1.51({ }3)(19) & \multicolumn{1}{c|}{---} & 0.05 & 1.52 \\
 \hline
 \multicolumn{2}{l}{{ }$\beta=14.0$} & & & & & & \\
 {\bf F} & 1.2302(2)(3) & 0.2236(1)(5) & -0.64({ }1)(43) & 2.25({ }6)(37) &
 \multicolumn{1}{c|}{---} & 0.09 & 0.68 \\
 {\bf G} & 1.2302(2)(2) & 0.2236(2)(3) & -2.5{ }({ }5)(25) & 1.63(12)(162) &
 -10({ }3)(13) & 0.004 & 0.68 \\
 {\bf H} & 1.2302(2)(1) & 0.2236(2)(1) & -1.8{ }({ }3)(14) & 1.70(12)(71) &
 -5({ }2)({ }7) & 0.005 & 0.68 \\
 {\bf J} & 1.2300(2)(1) & 0.2239(1)(1) & \multicolumn{1}{c|}{---} &
 2.65(35)(37) & \multicolumn{1}{c|}{---} & 0.01 & 1.35 \\
 \hline
 \multicolumn{2}{l}{{ }$\beta=20.0$} & & & & & & \\
 {\bf F} & 1.2321(2)(1) & 0.18151(4)(1) & -0.89({ }2)(42) & 2.36({ }5)(45) &
 \multicolumn{1}{c|}{---} & 0.12 & 0.75 \\
 {\bf G} & 1.2321(3)(2) & 0.1846{ }(4)(3) & -3.5{ }(46)(29) & 1.6{ }(59)(29) &
 -15(26)(17) & 0.03 & 0.75 \\
 {\bf H} & 1.2321(2)(2) & 0.1815{ }(1)(2) & -2.5{ }({ }2)(12) & 1.6{ }({ }1)(156) &
 -7({ }1)({ }8) & 0.03 & 0.75 \\
 {\bf J} & 1.2318(2)(2) & 0.1818{ }(1)(2) & \multicolumn{1}{c|}{---} &
 2.93({ }2)(17) & \multicolumn{1}{c|}{---} & 0.21 & 1.19 \\
 \hline
 \multicolumn{2}{l}{{ }$\beta=25.0$} & & & & & & \\
 {\bf F} & 1.2324(1)(2) & 0.1570(1)(2) & -0.92({ }4)(15) & 5.0{ }({ }4)(28) &
 \multicolumn{1}{c|}{---} & 0.03 & 0.70 \\
 {\bf G} & 1.2326(1)(3) & 0.1568(1)(2) & -2.8{ }({ }2)(15) & 1.7{ }({ }4)(272) &
 -10({ }1)(10) & 0.17 & 0.70 \\
 {\bf H} & 1.2325(1)(3) & 0.1568(1)(2) & -2.0{ }({ }1)(12) & 1.8(4)(214276) &
 -5({ }1)({ }5) & 0.16 & 0.70 \\
 {\bf J} & 1.2323(1)(1) & 0.1571(1)(1) & \multicolumn{1}{c|}{---} &
 2.85({ }4)(12) & \multicolumn{1}{c|}{---} & 0.22 & 1.28 \\
 \hline
 \multicolumn{2}{l}{{ }$\beta=31.0$} & & & & & & \\
 {\bf F} & 1.2326(1)(1) & 0.13530(2)(1) & -0.90({ }1)(20) & 2.55({ }1)(37) &
 \multicolumn{1}{c|}{---} & 0.50 & 0.74 \\
 {\bf G} & 1.2327(1)(1) & 0.13525(3)(1) & -3.0{ }({ }2)(11) & 1.67({ }5)(25) &
 -11({ }1)({ }7) & 0.31 & 0.74 \\
 {\bf H} & 1.2327(3)(4) & 0.13527(2)(2) & -2.2{ }({ }7)({ }6) & 2(335)(34) &
 5({ }7)({ }6) & 0.32 & 0.74 \\
 {\bf J} & 1.2324(1)(1) & 0.13543(3)(1) & \multicolumn{1}{c|}{---} &
 2.57({ }6)({ }9) & \multicolumn{1}{c|}{---} & 0.38 & 1.57 \\
 \hline
 \hline
\end{tabular}
\caption{Results of the fits for the extraction of \btt{} and $m$ for
$\SU(3)$ gauge theory.}
\label{tab:b2m-fits-su3}
\end{table}

The fit results are tabulated in tables~\ref{tab:b2m-fits-su2}
and~\ref{tab:b2m-fits-su3}. The first thing we note is that fit {\bf J} needs a
value of $R_{\rm min}$ which is a factor between 1.5 and 2 larger than $R_{\rm
min}$ for the other fits. In particular, $R_{\rm min}$ is typically larger than
those values of $R$ where we found the $R^{-4}$ correction term to be dominant
in section~\ref{sec:LO-correction}. This basically rules out the possibility
that the boundary term is absent, in which case the full $R^{-4}$ correction
would be given by the correction term from the massive modes. For the other fits
the values of $R_{\rm min}$ are comparable, but typically a bit smaller than
those of the fits in the previous section. This is not surprising given the fact
that each fit has an additional free parameter compared to fits {\bf B} to {\bf
D} from the previous section. When looking at the result for \btt{} and $m$ we
see that the fits {\bf F} to {\bf H} always agree within uncertainties, since
the uncertainties of the fits {\bf G} and {\bf H} are large. 
The large uncertainties are most likely due to the fact that the available
data does not allow to constrain the additional fit parameter in these
fits compared to fit {\bf F}. In the following analysis we will thus only
use the results from fit {\bf F}. The uncertainties of the parameters from
the other fits do not allow for any conclusions about the values of the
parameters and continuum extrapolations. In this case our results for the
individual lattice spacings are given by the values for fit {\bf F} in
tables~\ref{tab:b2m-fits-su2} and~\ref{tab:b2m-fits-su3} and they only
have two uncertainties, namely the statistical one and the systematic
uncertainty associated with the particular choice for $R_{\rm min}$
(estimated in the usual way). Unfortunately we cannot investigate the
systematic uncertainty due to unknown higher order correction terms, but
we believe that the presence of these terms is only a minor effect if
massive modes are present.

\begin{figure}[t]
 \centering
\includegraphics[]{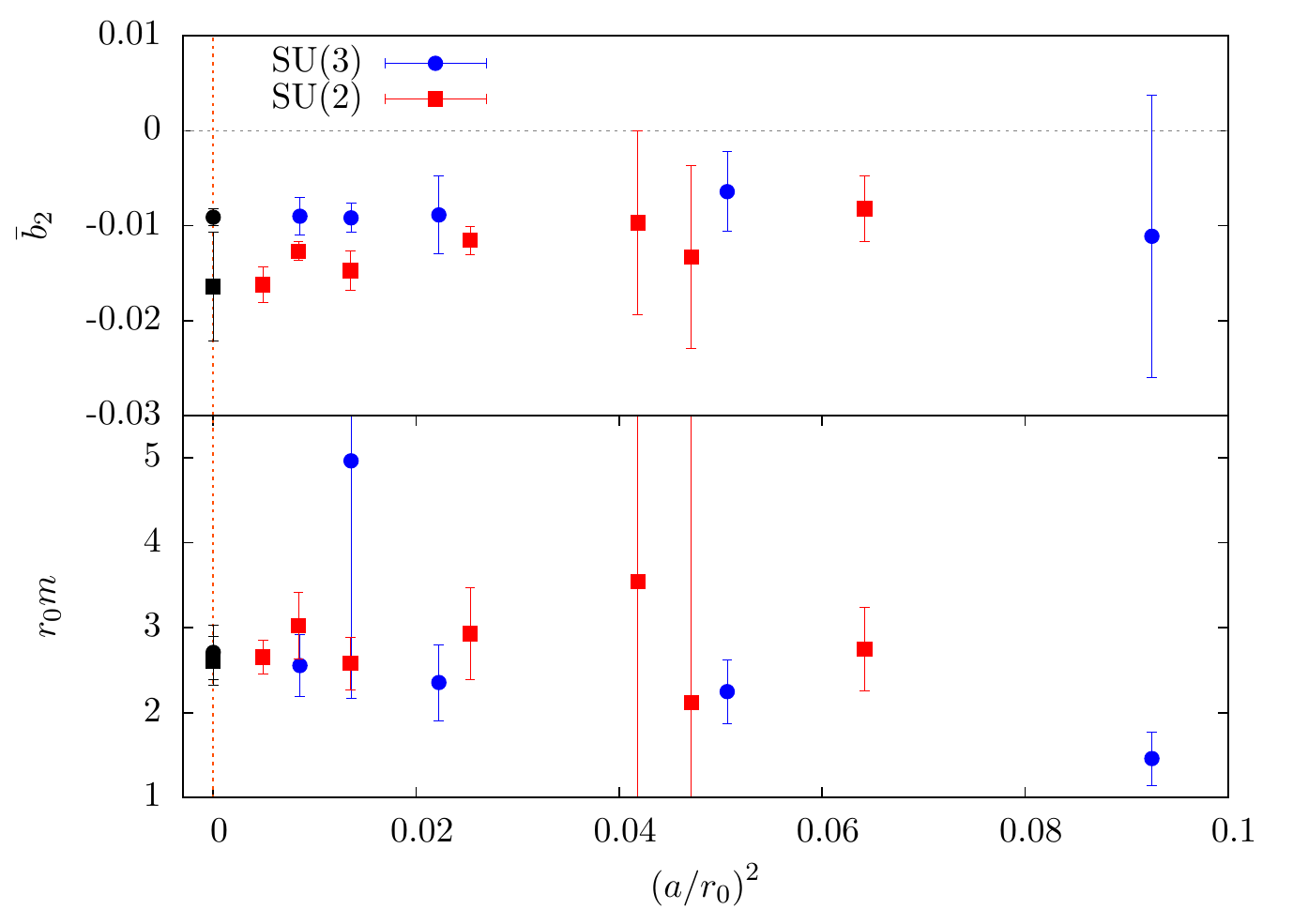}
 \caption{Results for \btt{} and $r_0m$ versus the  squared lattice
 spacing in units of $r_0$. Also shown are the continuum results from
 eqs.~\refc{eq:b2-conti-res2} and~\refc{eq:m-conti-res2}.}
 \label{fig:bm_vs_r0}
\end{figure}

The results for \btt{} and $r_0m$ are plotted versus $a^2$ in
figure~\ref{fig:bm_vs_r0}. In comparison to the results for \btt{} from the
previous section (figure~\ref{fig:br_vs_r0}), the results for $\SU(2)$ show
stronger fluctuations in the approach to the continuum while the results for
$\SU(3)$ obtain larger uncertainties. In general, the results for \btt{}
are closer to zero when massive modes are included. The reason for this
is obviously given by the presence of the additional $R^{-4}$ term in
eq.~\refc{eq:mass-fit}. This term contaminates the boundary correction
term and thus reduces the magnitude of \btt{}. The results for $r_0m$ show
a rather smooth approach to the continuum with the exception of the point
at $\beta=25.0$ for $\SU(3)$ gauge theory, which, however, also has a large
uncertainty.

\subsection{Continuum extrapolations}
\label{sec:b2m-conti}

Ultimately we are interested in the comparison between the two sets of
results in the continuum. To this end we perform a continuum extrapolation
similarly to the one in section~\ref{sec:b2-conti} by using fits of the
type (1) to (3) for \btt{} with the ansatz from eq.~\refc{eq:b2-conti}.
For $r_0m$ the ansatz reads
\be
\label{eq:m-conti}
r_0m = \big( r_0m \big)\cont + b_{m,1} \Big(\frac{a}{r_0}\Big)^2 + b_{m,2}
\Big(\frac{a}{r_0}\Big)^4
\ee
and we perform fits of the types (1) to (3) where $b_{\bt,2}$ is replaced
by $b_{m,2}$ in the fit definitions. The difference to
section~\ref{sec:b2-conti} is, that we only perform continuum extrapolations
for fit {\bf F}, so that no averaging in the continuum limit is necessary.
Once more the fits have also been done for the results from $R_{\rm min}\pm1a$
to assess the uncertainty associated with the choice for $R_{\rm min}$.

\begin{table}
\small
\centering
\begin{tabular}{cc|ccc}
 \hline
 \hline
 $N$ & Fit & (1) & (2) & (3) \\
 \hline
 2 & $\big( \bt \big)\cont$ & -0.0164(4)(56) & -0.0151(3)(27) & 
 -0.0164(5)(56) \\
 3 & $\big( \bt \big)\cont$ & -0.0087(3)({ }3) & -0.0098(1)(13) &
 -0.0091(2)({ }5) \\
 \hline
 2 & $\big( r_0m \big)\cont$ & 2.56({ }7)(31) & 2.65({ }3)(29) &
 2.61({ }6)(28) \\
 3 & $\big( r_0m \big)\cont$ & 2.75(12)(25) & 2.60({ }5)(62) &
 2.71({ }9)(29) \\
 \hline
 \hline
\end{tabular}
\caption{Results for $\big( \bt \big)\cont$ and $\big( r_0m \big)\cont$ from
fits (1) to (3) (see text). The first error is the statistical uncertainty,
the second the systematical error due to the choice for $R_{\rm min}$.}
\label{tab:b2m_conti_fitpar}
\end{table}

The results for the continuum extrapolations with fits (1) to (3) are given
in table~\ref{tab:b2m_conti_fitpar}. We usually obtain a rather good continuum
extrapolation, even though for \btt{} in $\SU(2)$ gauge theory the
extrapolation leads to rather large values of $\chi^2$/dof around 5 to 10, due
to the fluctuations of \btt{} for the small values of $a$. Since the continuum
extrapolation is a bit more problematic in this case we will use the results
from fit (3) for our final results. This fit appears to be a bit more stable
than the continuum extrapolation linear in $a^2$ using all data points. Once
more the uncertainty concerning the continuum extrapolation is estimated via the
maximal difference of the result of fit (3) with respect to the result of the
other fits. The continuum results for \btt{} and $r_0m$, given in
eqs.~\refc{eq:b2-conti-res2} and~\refc{eq:m-conti-res2} in
section~\ref{sec:results}, are shown in figure~\ref{fig:bm_vs_r0}. Owing to
the figure, we expect the hierarchy between the results from $\SU(2)$ and
$\SU(3)$ gauge theory to remain in the analysis with massive modes. The
inclusion of massive modes only changes the quantitative results while retaining
the qualitative features. The results for $r_0m$ in the continuum agree within
error bars between $\SU(2)$ and $\SU(3)$.

\section{Summary and discussion of the final results}
\label{sec:results}

The previous three sections contained a number of interesting results, which
are, however, difficult to extract from the somewhat lengthy discussion. Before
we move on to the conclusions let us thus summarise and discuss the main
findings in some more detail.

\subsection{Summary of results}
\label{sec:results-sum}

In section~\ref{sec:groundstate} we have extracted the Sommer
parameter and the string tension with high accuracy from our results for the
static potential. The results are given in table~\ref{tab:scale-fits}. After
that we have performed a continuum extrapolation of the string tension to obtain
the final result
\be
\label{eq:sig-conti-final}
\big( \sqrt{\sigma} r_0 \big)\cont = \left\{ \begin{array}{ll}
1.2356(3)(1) & \quad \text{for} \:\:\SU(2) \vspace*{2mm} \\
1.2325(3)(2) & \quad \text{for} \:\:\SU(3) \,.
\end{array} \right.
\ee
Here the string tension has been extracted by parameterising the potential
with the LC spectrum, eq.~\refc{eq:LC-spectrum}, up to the point
where the result for $\sigma$ did agree with the NLO expansion of
eq.~\refc{eq:LC-spectrum}. The first uncertainty is the statistical one
and the second uncertainty originates from the continuum extrapolation. In
three dimensions any comparison to physical units is meaningless. Nonetheless,
identifying $r_0$ with 0.5 fm~\cite{Sommer:1993ce} (to define the unit `fm'
in three dimensions) and using the standard conversion factor 
$\hbar c=197.3 (\ldots)$~fm MeV we obtain
\be
\label{eq:sig-conti-final-fm}
\big( \sqrt{\sigma} \big)\cont = \left\{ \begin{array}{ll}
487.6(2)(1) \:\:\text{MeV} & \quad \text{for} \:\:\SU(2) \vspace*{2mm} \\
486.4(2)(1) \:\:\text{MeV} & \quad \text{for} \:\:\SU(3) \,,
\end{array} \right.
\ee
respectively. A comparison to the latest results for the continuum string
tension in three dimensions~\cite{Lucini:2002wg,Bringoltz:2006gp} is only
possible in terms of the Karabili-Kim-Nair prediction~\cite{Karabali:1998yq},
i.e. in terms of the continuum extrapolated coupling. We leave this type of
comparison to the next publication. Our results at finite lattice spacing are
fully in agreement, but more precise, than the values given
in~\cite{Teper:1998te,HariDass:2007tx}. With a constant
continuum extrapolation the $\SU(2)$ results from~\cite{HariDass:2007tx}
give a continuum result around 1.234, which is a bit smaller but comparable to
our result.

After the extraction of the leading order (linear) behaviour we have
compared the results for the potential to the leading order EST prediction,
namely the LC spectrum, and extracted the exponent of the leading order
correction in $1/R$. The results, shown in figure~\ref{fig:m-exponent}
are consistent with an exponent of 4, as expected from the EST, where
the leading order correction is the boundary term proportional to \btt{}.
We then extracted the value for \btt{}, excluding contributions from
possible massive modes (cf. section~\ref{sec:beyondEST}). The results
at finite lattice spacing are given in table~\ref{tab:b2-results} and
the continuum results from the individual extrapolations in
table~\ref{tab:b2_conti_fitpar}. As the final continuum estimates for
\btt{} we obtain 
\be
\label{eq:b2-conti-res}
\big( \bt \big)\cont = \left\{ \begin{array}{ll}
-0.0257\,(3)(38)(17)(3) & \quad \text{for} \:\:\SU(2) \vspace*{2mm} \\
-0.0187\,(2)(13)(\:\:4)(2) & \quad \text{for} \:\:\SU(3) \,.
\end{array} \right.
\ee
The first error is purely statistical, the second is the systematical error
due to the unknown correction terms to the potential, the third is the
one associated with the particular choice for the minimal value of $R$
included in the fit for the extraction of \btt{}, $R_{\rm min}$, and the
fourth systematic uncertainty is the one due to the continuum extrapolation
(for details see section~\ref{sec:b2-ana1}). These values indicate that
the magnitude of \btt{} decreases with increasing $N$, meaning that it
could potentially vanish in the limit $N\to\infty$. The result also shows
that \btt{}, indeed, is non-universal, as expected since its value is not
constrained in the EST.

Concerning the extraction of \btt{} the main uncertainty comes from the
fact that massive modes, or, equivalently, a possible rigidity term in
the EST, lead to a contaminating additional $R^{-4}$ correction. The
presence of this term is expected to change the value of \btt{}. At
present it is unclear whether such a contamination will ultimately be present or
not, so that we have to take this possibility into account. On the basis of
our simulations we also cannot exclude this possibility, since the fits
to the potential reported in section~\ref{sec:m-test} work equally well
compared to the results from section~\ref{sec:b2-extract}. We can exclude,
however, the possibility that the $R^{-4}$ correction is fully due to the
correction associated with the massive modes, since fit {\bf J} (see
section~\ref{sec:m-test}) leads to a much larger value for $R_{\rm min}$ than
the other fits. Repeating the whole analysis, we see that the accuracy is only
sufficient for fit {\bf F} from tables~\ref{tab:b2m-fits-su2}
and~\ref{tab:b2m-fits-su3}. Using these results we obtain the continuum results
listed in table~\ref{tab:b2m_conti_fitpar} and the final continuum results
\be
\label{eq:b2-conti-res2}
\big( \bt \big)\cont = \left\{ \begin{array}{ll}
-0.0164\,(5)(56)(13) & \quad \text{for} \:\:\SU(2) \vspace*{2mm} \\
-0.0091\,(2)(\:\:5)(\:\:7) & \quad \text{for} \:\:\SU(3) \,.
\end{array} \right.
\ee
Here the first error is purely statistical, the second is the one
associated with the particular choice for $R_{\rm min}$ and the third
systematic uncertainty is the one due to the continuum extrapolation.
In this analysis we have only been able to use one of the fits for a
reliable extraction of \btt{} (fit {\bf F} from tables~\ref{tab:b2m-fits-su2}
and~\ref{tab:b2m-fits-su3}), which is why one of the systematic uncertainties
cannot be estimated. The comparison of eqs.~\refc{eq:b2-conti-res}
and~\refc{eq:b2-conti-res2} reveals that the inclusion of the contamination
reduces the magnitude of \btt{} but does not alter the qualitative feature of a
decrease in magnitude of \btt{} with increasing $N$.

From the extraction of \btt{} when massive modes are included we also obtain an
estimate for the mass of the massive modes. The
continuum results are listed in table~\ref{tab:b2m_conti_fitpar}, and as
our final continuum results we get
\be
\label{eq:m-conti-res2}
\big( r_0m \big)\cont = \left\{ \begin{array}{ll}
2.61\,(6)(28)(\:\:5) & \quad \text{for} \:\:\SU(2) \vspace*{2mm} \\
2.71\,(9)(29)(11) & \quad \text{for} \:\:\SU(3) \,,
\end{array} \right.
\ee
where the uncertainties are as in eq.~\refc{eq:b2-conti-res2}. The
interpretation of this ``mass'' is an open question. The first possibility is
that it represents a massive mode on the string worldsheet, which could indicate
that we observe the three-dimensional analogue to the worldsheet axion (cf.
section~\ref{sec:beyondEST}).
Recall, that the topological coupling term does not exist in 3d, so that
the massive mode will appear as a quasi-free mode on the worldsheet up to
coupling terms including two massive boson fields. On the
other hand, the contribution could also be due to the rigidity term, in which
case the results from eq.~\refc{eq:m-conti-res2} can be translated into results
for the coupling $\alpha$ via eq.~\refc{eq:mdef}. It is intriguing to note
that the results for $m$, dividing by $\sqrt{\sigma}$ from
eq.~\refc{eq:sig-conti-final}, are very similar to the results for the mass of
the worldsheet axion from~\cite{Dubovsky:2013gi,Athenodorou:2017cmw}, at least
for the case of $\SU(2)$ gauge theory. For $\SU(3)$ gauge theory our result is
somewhat larger than the result
from~\cite{Dubovsky:2013gi,Athenodorou:2017cmw}, however, in contrast to
those results our result is extrapolated to the continuum. In fact, our
results around $\beta=25.0$ are already fully compatible with the results
from~\cite{Dubovsky:2013gi,Athenodorou:2017cmw}. This opens up the possibility
that, indeed, we are looking at a massive mode on the three dimensional
worldsheet which is similar in nature to the worldsheet axion.

It is also important to ask whether the value for $m$ extracted in our
analysis does still comply with the framework of the EST. From
eq.~\refc{eq:break-scale} we expect modes of masses down to a few times
$\sqrt{\sigma}$ to be integrated out. The masses we have obtained here are
around twice as large as $\sqrt{\sigma}$. In our fits we could typically go
down to $R/r_0\approx0.7$. When we assume that this is a sign for the scale
where the EST is bound to break down, this leads to a cut-off scale of
$1.2\times\sqrt{\sigma}$. However, this estimate could well underestimate the
true cut-off scale, so that a mass of $2\times\sqrt{\sigma}$ can potentially be
below that bound.

\subsection{Comparison between continuum EST and data}

\begin{figure}[t]
 \centering
\includegraphics[]{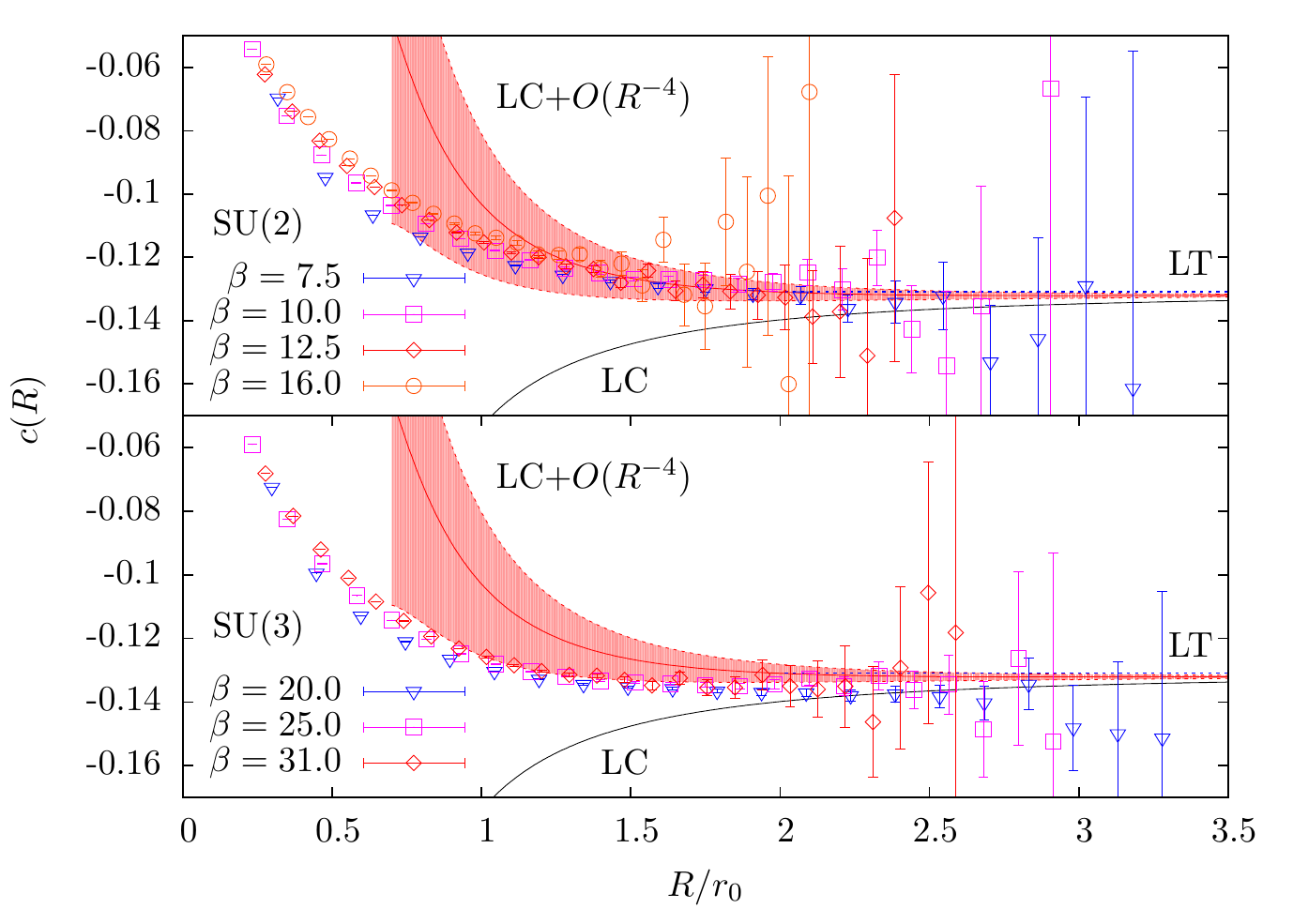}
 \caption{Results for the curvature $c(R)$ plotted versus $R/r_0$. The curve
labeled with `LT' is the L\"uscher constant, the one labeled with `LC' is the
light cone potential with the continuum extrapolated string tension and the
curve labeled with `LC$+O(R^{-4})$' is the LC curve including the boundary term
with the continuum extrapolated value of \btt.}
 \label{fig:cr-coupling}
\end{figure}

We will now compare the continuum predictions for the potential from the EST to
the results for the potential on our finest lattice spacings. A suitable
quantity to visualise the subleading contributions to the potential is the
curvature, associated with the second derivative
\be
\label{eq:cr-term}
c(R) = \frac{R^3}{2} \frac{\partial^2 V(R)}{\partial R^2} \,.
\ee
In the $R\to\infty$ limit $c(r)$ is expected to reproduce the L\"uscher constant
$-\pi/24$. The results for $c(R)$ are compared to the continuum results for the
EST in figure~\ref{fig:cr-coupling}. The black curve in the figure is the LC
spectrum, eq.~\refc{eq:LC-spectrum}, with the continuum string tensions from
eq.~\refc{eq:sig-conti-final} (the uncertainties are smaller than the line), and
the red band includes the boundary term from eq.~\refc{eq:est-spec} with the
continuum boundary coefficient \btt{} from eq.~\refc{eq:b2-conti-res}. The
inclusion of the boundary term obviously enhances the agreement with the data
significantly down to small values of $R$. We also note that the two curves
are basically indistinguishable at the scale of
the plot (they are not identical, however). For $\SU(2)$ gauge theory the data
at finite lattice spacing already agrees rather well with the continuum EST,
while for $\SU(3)$ gauge theory the data lies below the curve. We have not shown
the curve including massive modes in the plot since it overlaps with the
`LC$+O(R^{-4})$' curve and compares very similarly to the data.

\section{Conclusions}
\label{sec:concl}

In this article we have performed a high precision study of the static
$q\bar{q}$ potential in three-dimensional $\SU(N)$ gauge theory with $N=2$ and
3 and compared the results to the potential obtained from the effective string
theory. In particular, we obtained accurate results with full control over the
systematic effects for the continuum string tension, the non-universal boundary
coefficient \btt{} and, for the extended analysis, the ``mass'' of the possible
massive mode within the EST. The results are summarised and discussed in detail
in section~\ref{sec:results-sum}. In particular, we could show that the leading
order correction to the light cone spectrum is of $O(R^{-4})$ (cf.
section~\ref{sec:LO-correction}). If massive modes are present, the results
for \btt{} change (see eqs.~\refc{eq:b2-conti-res} and~\refc{eq:b2-conti-res2})
due to another correction of $O(R^{-4})$, contaminating the result for \btt{}.
However, the contribution from the massive modes only is not enough to describe
the data, so that we can conclude that $\bt\neq0$.~\footnote{This statement is
true up to the unlikely possibility of fine-tuned higher order corrections,
mimicking the effect of an $O(R^{-4})$ term.}

The data for \btt{} shows an interesting trend towards zero with increasing
$N$, leading to the possibility that \btt{} could vanish in the limit
$N\to\infty$. This result is also independent of whether or not we include
massive modes in the analysis. In the context of generalisations of the AdS/CFT
correspondence to large $N$ gauge theories, this is an interesting result
since it imposes constraints on the fundamental string theory for the dual
version of Yang-Mills theories at large $N$. We will investigate this issue
further in the next publication in this series of papers.

The main obstacle of our analysis for \btt{} is the fact, that it is impossible
to judge whether or not massive modes, or, equivalently, the rigidity term
(or even both) are present. While some properties (like
the decrease in magnitude for $N\to\infty$) are independent of this issue, the
exact value of \btt{} can only be extracted once the issue is resolved. To this
end it would be desirable to have an expression for the excited state energies
in the presence of a massive mode, or, alternatively, the rigidity term. This
could potentially help to discriminate between the existence or non-existence
of such terms and, in addition, it could help to discriminate between the two
types of additional contributions.

It is intriguing to see that the results for the mass of the possible massive
modes is in good agreement with the masses found in
~\cite{Dubovsky:2013gi,Athenodorou:2017cmw}. It is thus possible that we are
seeing a massive mode on the worldsheet which is similar in nature to the
worldsheet axion. Note, that in 3d the topological coupling term is absent,
so that the analogue to the worldsheet axion appears as a quasi-free mode
on the worldsheet coupling to the GBs via the covariant derivative and possible
higher order coupling terms. It will be interesting to see whether the mass of
the massive modes remains consistent with the one of the worldsheet axion
for $N\to\infty$.

\section*{Acknowledgements}

I am grateful to Marco Meineri for numerous enlightening discussions during
the collaboration on the joint review~\cite{Brandt:2016xsp} and I would like to
thank F. Cuteri for carefully going through the manuscript. The simulations have
been done in parts on the clusters Athene and iDataCool at the University of
Regensburg and the FUCHS cluster at the Center for Scientific Computing,
University of Frankfurt. I am indepted to the institutes for offering these
facilities. During this work I have received support from DFG via SFB/TRR 55 and
the Emmy Noether Programme EN 1064/2-1.

\appendix

\section{Simulation setup}
\label{app:sim-setup}

This appendix contains the details of the numerical simulations. For the
configuration updates we have used the, Cabbibo-Marinari heatbath
algorithm~\cite{Cabibbo:1982zn} for $\SU(N)$ gauge theories with all $\SU(2)$
subgroups, in combination with the improved $\SU(2)$ heatbath algorithm
from~\cite{Kennedy:1985nu}. For each update we perform three overrelaxation
steps~\cite{Creutz:1987xi}.

To obtain a clean result for the groundstate potential in terms of
contaminations from excited states the spatial correlation function of two
Polyakov loops is the most promising observable. Its spectral representation is
given by
\be
\label{eq:pcor}
\ew{P^{\ast}(R)\:P(0)} = \sum_{i=0}^{\infty} b_i \: e^{-E_i(R)\:T}
\ee
for $R<L/2$, where $T=a\;N_t$ and $L=a\;N_s$ are the temporal and spatial
lengths of the lattice, $b_i$ is the overlap between the operator and the
associated energy eigenstate and $E_i(R)$ is the $i$'th energy level. We always
take energies to be ordered in ascending order, i.e. $E_0<E_1<E_2<\ldots\;$. In
the limit of $T\to\infty$ the leading contribution to eq.~\refc{eq:pcor} is
given by the groundstate $E_0(R)=V(R)$ and excited states are suppressed with
$\exp\{-[E_i(R)-E_0(R)]\;T\}$. For large values of $T$ excited states can thus
be neglected to a good approximation and we check in
appendix~\ref{app:sys-effects} whether this is true for the temporal extents
used in the present study. If we can neglect contaminations from excited states
we can extract $V(R)$ via
\be
\label{eq:potential}
V(R) = -\frac{1}{T} \ln\left[\ew{P^{\ast}(R)\:P(0)}\right] \;.
\ee
Note, that the potential is determined from the effective string theory up to a
constant, which we denote as $V_0$ (it can also be related to the constant term
$\mu$ in the boundary action~\refc{eq:bound_action}; here, however, we are not
really interested in this quantity).

The string tension can either be extracted from a fit to the potential, or via
the force
\be
\label{eq:force}
F(R) \equiv \frac{\partial \; V(R)}{\partial R} \;,
\ee
given in the $1/R$ expansion of the EST in three dimensions by
\be
\label{eq:force_nlo}
F(R) = \sigma + \frac{\pi}{24} \; \frac{1}{R^2} + \Ord(R^{-4}) \,.
\ee
Note, that the determination via the force does not demand the determination of
$V_0$. Here we will use both methods to extract the string tension and
compare the results.

The string tension can also be used to fix the lattice spacing in physical
units and observables can then be expressed in units of $a\sqrt{\sigma}$.
However, the determination of $a\sqrt{\sigma}$ demands to control the asymptotic
$R\to\infty$ behaviour. A more suitable observable to set the reference scale is
the Sommer parameter $r_0$~\cite{Sommer:1993ce}, which is defined implicitly by
\be
\label{eq:sommer_para}
r_0^2\;F(r_0) = 1.65 \;.
\ee
$r_0$ can be obtained directly by interpolation of the results for $R^2\;F(R)$.

The reliable extraction of the potential demands the control of the
contaminations from excited states and thus large temporal extents. Since the
signal-to-noise ratio of such Polyakov loop correlation functions decays
exponentially with $T$, the reliable extraction of the expectation values
demands the use of a suitable error reduction algorithm. Our choice is to use
the multilevel algorithm introduced by L\"uscher and
Weisz~\cite{Luscher:2001up}. In particular, we use one level of averaging for
all lattices and the parameters such as the number of sublattice updates, $n_t$,
and the temporal extent of the sublattices, $t_s$, are given in
table~\ref{tab:sim-paras}.

\section{Control of systematic uncertainties}
\label{app:sys-effects}

For the measurements of the potential there are several systematic effects that
need to be controlled for a reliable extraction of the potential at a given
lattice spacing. In particular, this concerns the contaminations from excited
states and finite size effects. Checks concerning these two effects are
reported in the following. There is yet another effect at very fine lattice
spacings, owing to the ``freezing'' of the topological charge in certain
sectors. We control this effect by using a completely new run (including
full thermalisation starting from a random configuration) for at least each two
measurements. This can be done since the thermalisation (we take a very
conservatively long thermalisation with at least 1000 update steps) needs
relatively little time compared to one measurement with the multilevel
algorithm.

\subsubsection*{Excited state contributions}

\begin{figure}[t]
 \centering
\includegraphics[]{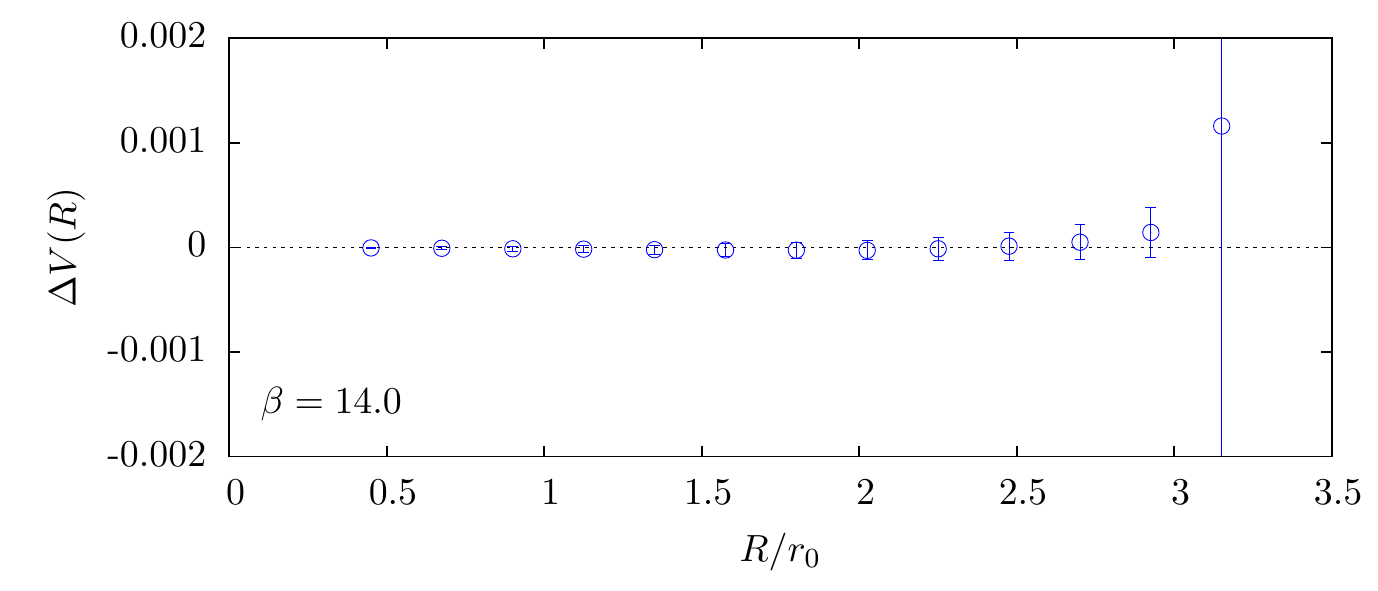}
 \caption{Difference in the results for the static potential $\Delta V$ from
temporal extents 48 and 64 at $\beta=14.0$ versus the $q\bar{q}$-separation in
units of $r_0$.}
 \label{fig:excited_states_test}
\end{figure}

Following the spectral representation~\refc{eq:pcor} for the Polyakov loop
correlation function, the correction to eq.~\refc{eq:potential} owing to excited
states is given by (see also~\cite{Luscher:2002qv})
\be
\label{eq:PP_exc_contr}
V(R) = -\frac{1}{T} \ln\left[\ew{P^{\ast}(R)\:P(0)} + w_1 e^{-\Delta E \,T} +
\ldots \right] \;.
\ee
The contamination from the second term on the right hand side can be checked by
performing simulations for two different temporal extents $T$ of the lattice.
Here we have done simulations for $\SU(3)$ with $\beta=14.0$ on 48 and
$64\times48^2$ lattices and compare the results for the potential. The results
for the difference of the two energies are shown in
figure~\ref{fig:excited_states_test}. As can be seen from the plot,
the two sets of energies agree perfectly within errors for the whole range of
$q\bar{q}$ separations considered, meaning that excited states are negligible.
Since for the different ensembles $T$ changes only little in physical units, we
expect this to hold for all lattices considered. Note, that the energy levels
(and the overlaps) of the flux tube, in general, change only on the subleading
level for different values of $N$. Consequently, we would expect this result to
be valid for different values of $N$ as well.

\subsubsection*{Finite size effects}

For Polyakov loop correlators there are two types of finite size effects that can occur:
(a) the flux tube feels the extent of the finite lattice volume perpendicular to its extension;
(b) the contribution to the energies of the flux tube winding ``around the world''.
Relevant for effect (a) is the bulk correlation length and, with our lattice
volumes, this effect should be suppressed by many orders of magnitude. Effect
(b) might cause trouble and its effect should be investigated. Including the
winding contribution, the Polyakov loop correlator is to leading order given by
\be
\label{eq:PP-enes-fse}
\ew{P^{\ast}(R)\:P(0)} = b_0 \: \left[ e^{-E_0(R)\:T} + e^{-E_0(L-R)\:T}
\right] \,.
\ee
Since $E_0(R)$ growths linearly with $R$, we expect the winding contribution to
be strongly suppressed when $L-R>R$.

Assuming the leading order linear relation $\sigma R$ for the bare energy in
lattice units and a relative error of $10^{-3}$ for the Polyakov loop correlator
(which is a lower bound for what we get for the larger $R$ values), we see that
the second term is negligible if
\be
\label{eq:fse-criterion}
\frac{1}{2} \Big( \sqrt{\sigma} L - \frac{\ln(3)}{\sqrt{\sigma}T} \Big) >
\sqrt{\sigma}R \quad \Rightarrow \quad \sqrt{\sigma}R < 6.1 \,,
\ee
where in the second step we have used that for our lattices $T=L$ and
$\sqrt{\sigma}L\gtrsim12.3$. This condition is always fulfilled in our study.

\begin{figure}[t]
 \centering
\includegraphics[]{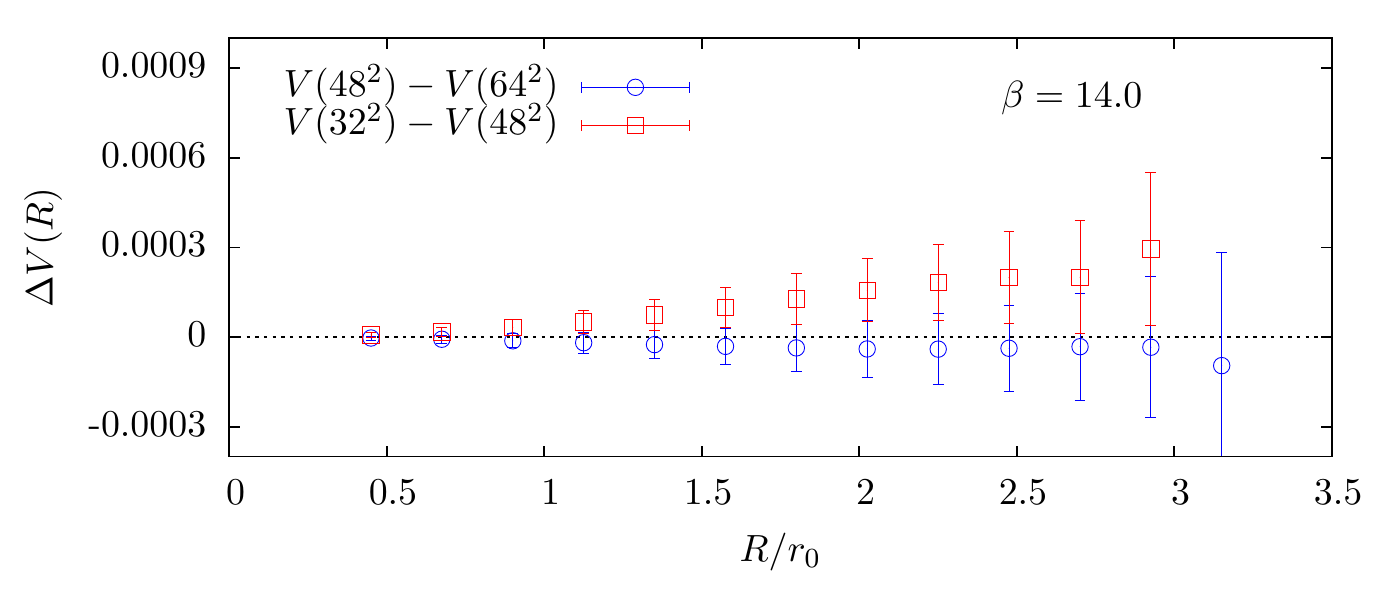}
 \caption{Difference in the results for the static potential $\Delta V$ from
volumes $48^2$ and $64^2$ (blue circles) and $32^2$ and $48^2$ with $N_t=48$
versus the $q\bar{q}$-separation in units of $r_0$ at a temporal extent of 48
and $\beta=14.0$.}
 \label{fig:FS_states_test}
\end{figure}

As an additional check we computed the potential on a $48\times64^2$ lattice
for $\SU(3)$ with $\beta=14.0$ and compared the results for the potential with
the $48^3$ run used in the analysis. The results are shown in
figure~\ref{fig:FS_states_test}. The plot indicates that the effects due to the
finite lattice extent are negligible. Note, however, that finite size effects
can be present for these values of $R$ if the lattice is chosen to be smaller.
This can be seen from the comparison to the results for a $48\times32^2$
lattice, also shown in figure~\ref{fig:FS_states_test}.

We also note that another finite size effect could be due to a glueball
exchange between the two Polyakov loops. Generically we expect this effect to
be suppressed by a factor of $\exp(-m_G R)$, or $\exp(-m_G(L-R))$. Clearly, as
long as $m_G>\sqrt{\sigma}$ (which is always the case) this contribution
appears as an excited state effect, which we have already ruled out in the
check for excited state contaminations.

\bibliographystyle{JHEP}
\bibliography{paper}

\end{document}